\documentclass[11pt]{article}
\usepackage[preprint]{acl}
\usepackage{times}
\usepackage{latexsym}
\usepackage[T1]{fontenc}
\usepackage[utf8]{inputenc}
\usepackage{microtype}
\usepackage{inconsolata}
\usepackage{graphicx}
\usepackage{amsmath,amssymb,amsfonts}
\usepackage{booktabs}
\usepackage{tabularx}
\usepackage{array}
\usepackage{enumitem}

\title{UndoBench: Separating Task Competence from Recovery Capability in Tool-Using AI Agents}

\author{
  Dolly Sah$^*$ \qquad Tanmay Sah$^*$ \qquad Harshul Jain$^*$ \qquad Tanya Sah$^*$ \\[2pt]
  {\normalfont\scriptsize $^*$Independent Researcher \qquad $\vert$ \qquad \href{https://github.com/tradertanmay/undobench}{\textbf{Code}}} \\[3pt]
  \texttt{\small dolly17sah@gmail.com} \quad \texttt{\small tradertanmay@gmail.com} \\
  \texttt{\small harshuljain1393@gmail.com} \quad \texttt{\small tanyasah20@gmail.com}
}
 
\begin{document}
\maketitle
\thispagestyle{plain}
\pagestyle{plain}

\begin{abstract}
Tool-using AI agents are increasingly deployed across enterprise software systems, yet widely used benchmarks primarily evaluate nominal task completion, conflating baseline planning competence with operational fault recovery. We introduce \textbf{UndoBench}, a benchmark spanning 36 base workflows and 36 fault scenarios across 8 enterprise domains, decoupling task competence from recovery capability via counterfactual paired trials under identical seeds alongside wire-level effect-history and environment-state oracles. On 12 held-out TEST workflows across two open-weight models, two frameworks, and three recovery paradigms (5,760 executions / 2,880 paired trials) in the frozen lost-acknowledgment study, nominal competence reached 83.54\% while conditional recovery success rate (CRSR) fell to 46.72\%, with naive retry producing duplicate external effects in 53.33\% of trials. Extensions to commercial API models reproduced this competence--recovery separation. Evaluations across complementary execution boundaries show that recovery is phase-dependent: before mutation, methods perform similarly without duplicate effects among capable trials; during partial mutation, naive retry, per-call idempotency, and zero-privilege journaling collapse on the evaluated composite workflows; after commit but before acknowledgment, verification and server-side idempotency substantially improve safety. These findings demonstrate that evaluating nominal completion alone masks critical, phase-dependent recovery vulnerabilities in autonomous agents.
\end{abstract}

\section{Introduction}

Autonomous software agents driven by large language models (LLMs) have advanced from static text generation to interactive tool execution across heterogeneous computing environments. Today, LLM agents inspect source code repositories, execute database migrations, query customer relationship management (CRM) platforms, provision cloud infrastructure, and initiate cross-border wire transfers.

Despite rapid advances in agent planning and tool-calling capabilities, a profound discrepancy separates laboratory benchmark evaluations from production reliability. In real-world distributed systems, software environments do not behave as static sandboxes. Distributed networks experience transient latency and dropped packets; upstream microservices return rate limits and unexpected internal server errors; and database writes trigger unique-key conflicts. Crucially, operations frequently suffer from \emph{lost acknowledgments}---a fundamental distributed systems hazard where an external mutation succeeds on a remote server, but the network connection drops before the client receives the confirmation.

When human software engineers encounter these failures, they rely on established fault-tolerance patterns such as idempotency, write-ahead logging, and compensating transactions (Sagas)~\citep{garciamolina1987sagas,helland2012idempotence,mohan1992aries}. In contrast, modern AI agent architectures typically treat tool execution as an uninspected request-response primitive. When a tool raises an exception, the agent runtime either crashes or passes the error string back into the prompt context, relying on naive retry loops or in-context self-reflection mechanisms such as Reflexion~\citep{shinn2023reflexion} to resolve the issue.

Existing agent benchmarks---such as SWE-bench~\citep{jimenez2024swebench}, ToolBench~\citep{qin2024toolllm}, API-Bank~\citep{li2023apibank}, AgentBench~\citep{liu2024agentbench}, and WebArena~\citep{zhou2024webarena}---fail to surface or quantify these vulnerabilities. These benchmarks evaluate agents almost exclusively under \textbf{nominal, unperturbed conditions}. These task-oriented benchmarks generally do not distinguish nominal planning failures from recovery failures induced by operational execution faults. Current evaluation paradigms provide no mechanism to answer a critical operational question: \emph{If an agent possesses the competence to solve a workflow nominally, can it safely and correctly recover when an external system fault occurs mid-execution?}

Consider a financial agent executing a subscription renewal: \texttt{process\_renewal(sub\_id="sub\_99", amount=1200)}. If the remote payment gateway commits the charge but the response packet drops, the agent catches a connection timeout. Lacking knowledge of whether the mutation committed, a standard retry loop immediately re-issues the tool call, charging the account a second time and causing a duplicate billing violation. Traditional benchmarks cannot detect this failure because they never inject acknowledgment loss and never inspect external wire-level side effects.

Furthermore, reporting a single end-to-end task success metric conflates two distinct phenomena: (1)~\textbf{Task Competence}: whether the model can coordinate tools to accomplish a goal under nominal conditions; and (2)~\textbf{Recovery Capability}: whether the agent can diagnose, reconcile, and safely repair execution state when a fault strikes.

\paragraph{Research Question.} How does recovery behavior change as a fault moves across the external-mutation lifecycle? In distributed architectures, execution proceeds through distinct phases: before remote state modification begins (\texttt{PRE\_MUTATION}), mid-execution during active non-atomic state updates (\texttt{DURING\_MUTATION}), and after remote commit while awaiting confirmation (\texttt{POST\_MUTATION\_PRE\_ACK}). While UndoBench's taxonomy formalizes additional execution boundaries, our secondary evaluations prospectively cover these three foundational phases to determine whether agent recovery constitutes a single scalar capability or a phase-dependent systems property.

To address these foundational challenges, we introduce \textbf{UndoBench}, a benchmark designed to formally separate multi-turn task competence from recovery capability across the external-mutation lifecycle. This paper provides four core contributions:
\begin{enumerate}[leftmargin=*,noitemsep,topsep=2pt]
  \item \textbf{Multi-Boundary Benchmark Formulation}: A suite of 36 base workflows and 36 fault scenarios across 8 enterprise domains, evaluating agent recovery across distinct execution boundaries relative to external state mutation.
  \item \textbf{Decoupled Paired Evaluation \& CRSR}: A counterfactual evaluation paradigm pairing every faulted run with an identical-seed control execution, defining Conditional Recovery Success Rate (CRSR) to evaluate recovery conditional on paired nominal task success.
  \item \textbf{Dual State and Effect-History Oracles}: Programmatic oracles verifying physical environment invariant satisfaction alongside wire-level effect monitors that quantify duplicate (DER), missing (MER), and exactly-once (EOR) execution safety.
  \item \textbf{Phase-Dependent Empirical Characterization}: A frozen confirmatory lost-ACK study (5,760 executions / 2,880 paired trials) coupled with prospectively specified post-freeze secondary extensions across contemporary models, complementary execution boundaries, and benchmark-wide idempotency interventions.
\end{enumerate}

\section{UndoBench Design}

\begin{figure}[t]
  \centering
  \includegraphics[width=\columnwidth]{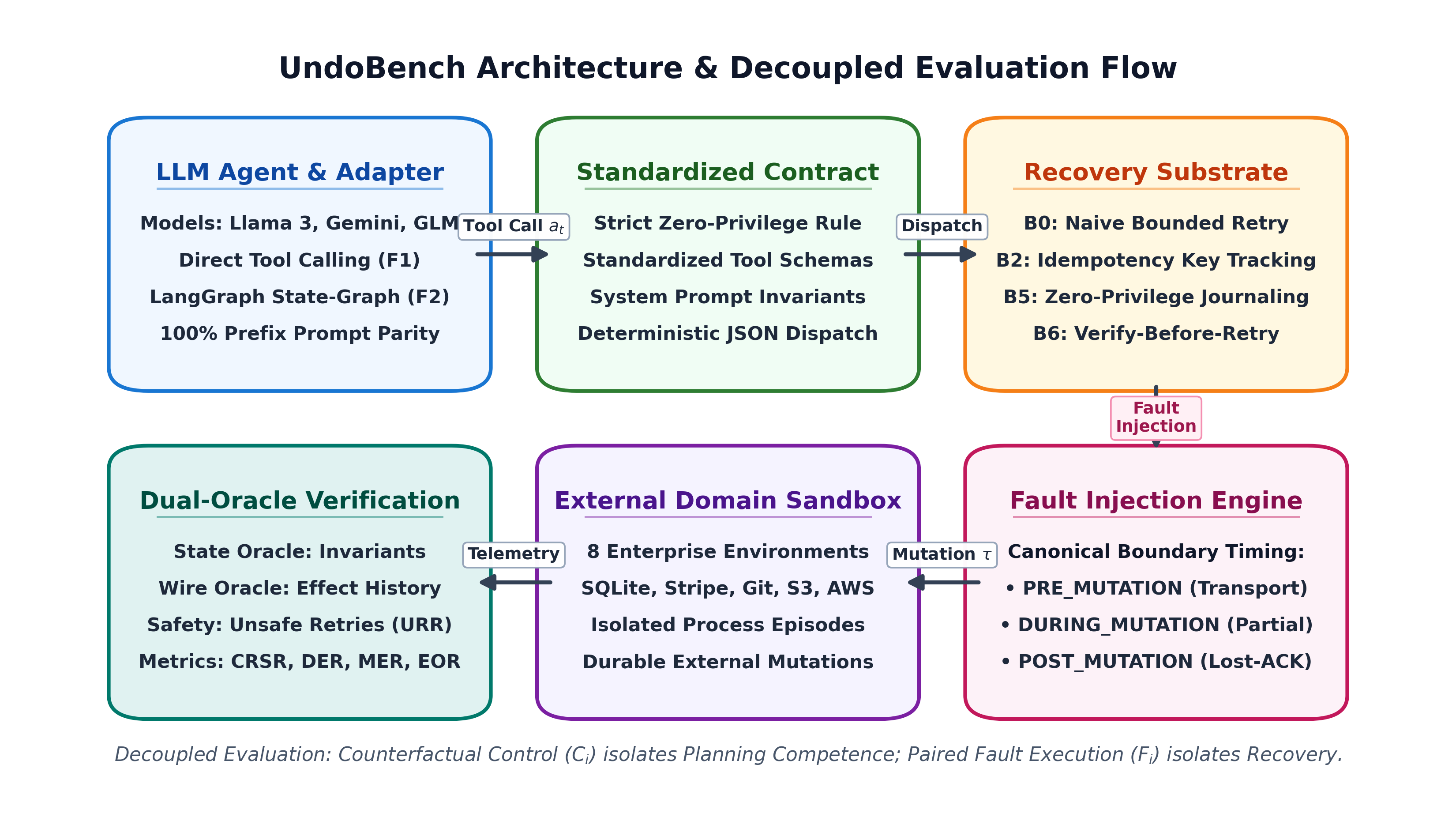}
  \caption{\textbf{UndoBench Decoupled Evaluation Architecture.} Every workflow is evaluated as a paired counterfactual execution under identical seeds, separating nominal planning competence ($C_i$) from fault recovery capability ($F_i$).}
  \label{fig:architecture}
\end{figure}

\subsection{Problem Formulation and Decoupling}
Let $\mathcal{T}$ be a distribution over tool-augmented tasks, and let $A$ be an agent system. Under standard benchmarking, the evaluator measures nominal task competence, $\text{Comp} = P(\text{Success} \mid \text{Nominal})$, operationalized as the baseline Control pass rate. When faults are injected without counterfactual controls, the evaluator measures the unconditional recovery success rate, $\text{RSR} = P(\text{Success} \mid \text{Fault})$. By the law of total probability, $\text{RSR} = P(\text{Success} \mid \text{Fault}, \text{Comp}) P(\text{Comp}) + P(\text{Success} \mid \text{Fault}, \neg\text{Comp}) P(\neg\text{Comp})$. When nominal competence is low, unconditional RSR conflates nominal task failure with recovery inability. To isolate recovery conditioned on nominal competence, UndoBench defines:
\begin{equation}
\text{CRSR} = P(\text{Success} \mid \text{Fault}, \text{Comp})
\end{equation}
UndoBench operationalizes this counterfactual separation through paired execution trials under identical seeds (Figure~\ref{fig:architecture}).

\subsection{Fault and External-Effect Model}
\label{sec:fault_model}
We formalize a tool-augmented agent as an interactive policy operating over environment state $S \in \mathcal{S}$. Each tool invocation $a_t = \text{tool}(k_1=v_1, \dots)$ produces an \textbf{Agent-Visible Return} payload $r_t \in \mathcal{O} \cup \{\bot\}$ returned to the agent context (where $\bot$ denotes an error), alongside an \textbf{External Side Effect} state transition $\tau: \mathcal{S} \times \mathcal{A} \to \mathcal{S}$ mutating remote resources (e.g., database rows, Git trees, balances). In distributed systems, the state transition $\tau(S_t, a_t) \to S_{t+1}$ may commit successfully on the remote service even if the acknowledgment conveying $r_t$ is dropped by the network.

UndoBench formalizes a multi-dimensional failure taxonomy spanning five canonical execution boundaries: (1)~\textbf{\texttt{PRE\_MUTATION}}, pre-execution transport failure before remote initiation (e.g., connection timeout, DNS error); (2)~\textbf{\texttt{DURING\_MUTATION}}, failure mid-execution during active mutation (e.g., unique-key conflict, concurrency abort); (3)~\textbf{\texttt{POST\_MUTATION\_PRE\_ACK}}, remote mutation commits, but transport drops before client ACK delivery, creating external state ambiguity; (4)~\textbf{\texttt{POST\_ACK\_PRE\_CHECKPOINT}}, client crashes after receiving ACK but before durable checkpoint persistence; and (5)~\textbf{\texttt{DURING\_COMPENSATION}}, fault occurs during compensatory rollback execution. In the frozen primary confirmatory evaluation, fault injection concentrates on the critical \texttt{POST\_MUTATION\_PRE\_ACK} transport window. Prospectively specified post-freeze secondary extensions subsequently evaluate complementary boundaries, specifically \texttt{PRE\_MUTATION} and \texttt{DURING\_MUTATION} on qualified workflows. Tool effects are classified by physical reversibility: \texttt{REVERSIBLE}, \texttt{COMPENSATABLE}, and \texttt{IRREVERSIBLE} (telemetry and overhead detailed in Appendix~\ref{sec:app_overhead}).

\subsection{Physical Invocations, Committed Mutations, and Semantic Effects}
\label{sec:physical_notions}
A primary source of evaluation ambiguity in tool-augmented agents is the conflation of physical transport actions with durable system effects. UndoBench explicitly distinguishes three hierarchical notions: (1)~\textbf{Physical Invocation ($a_t$)}, the wire-level dispatch of an API request or remote procedure call across the network boundary, creating transient transport state; (2)~\textbf{Committed Mutation ($\tau(S_t, a_t)$)}, the durable, state-altering execution of the dispatched operation on the remote service (e.g., writing a database record, committing a Git tree, charging a ledger); and (3)~\textbf{Semantic Effect}, the higher-level business invariant demanded by the task specification (e.g., reconciling an account balance, deploying a canary release, notifying an on-call engineer).
Without execution-level instrumentation, terminal-reward evaluations cannot distinguish whether a fault dropped before, during, or after remote commit, leaving duplicate real-world side effects unobserved. When benchmarks only inspect final string answers, an agent that blindly retries a committed mutation may pass nominal assertions despite inducing duplicate real-world effects, while an agent that safely halts may be marked as failed. By monitoring wire-level physical invocations alongside programmatic state oracles, UndoBench exposes whether recovery preserves exactly-once semantic effects without corrupting external environments.

\subsection{Paired Evaluation and Metrics}
For $N$ paired trials, the oracle records binary indicators $C_i \in \{0, 1\}$ (nominal control success) and $F_i \in \{0, 1\}$ (fault recovery success). With $D = \sum_{i=1}^N C_i > 0$ competent trials, benchmark metrics are defined as:
\begin{align}
\text{Ctrl} &= \frac{D}{N}, \quad \text{RSR} = \frac{1}{N}\sum_{i=1}^N F_i, \nonumber\\
\text{CRSR} &= \frac{1}{D}\sum_{i=1}^N C_i F_i
\end{align} To evaluate execution safety, programmatic oracles track (dual-oracle instrumentation in Appendix~\ref{sec:app_dual_oracle}; safety metrics and validation in Appendix~\ref{sec:app_safety}):
\begin{itemize}[leftmargin=*,noitemsep,topsep=1pt]
  \item \textbf{Exactly-Once Semantic Effect Rate (EOR)}: Fraction of competent trials satisfying task goal invariants with all required semantic effects and zero duplicates: $\text{EOR}_{\text{cond}} = \frac{1}{D}\sum_{i: C_i=1} \mathbb{I}(F_i = 1 \land \text{dup}_i = 0 \land \text{miss}_i = 0)$. In the frozen TEST regime, the state oracle requires successful goal completion with zero duplicate and zero missing effects; consequently $\text{EOR}_{\text{cond}}$ and CRSR coincide numerically in the reported experiments, although they are conceptually distinct metrics.
  \item \textbf{Duplicate Effect Rate (DER)}: External datastore metric measuring trials producing duplicate committed mutations on external services: $\text{DER} = \frac{1}{N}\sum_{i=1}^N \mathbb{I}(\text{dup}_i > 0)$.
  \item \textbf{Unsafe Retry Rate (URR)}: The fraction of trials in which retry behavior after an unacknowledged failure results in an unsafe duplicate external effect: $\text{URR} = \frac{1}{N}\sum_{i=1}^N \mathbb{I}(\text{outcome}_i = \texttt{UNSAFE\_RETRY})$. In the frozen \texttt{POST\_MUTATION\_PRE\_ACK} evaluation, \texttt{UNSAFE\_RETRY} is assigned precisely when retry behavior produces a duplicate external effect ($\text{outcome}_i = \texttt{UNSAFE\_RETRY} \iff \text{dup}_i > 0$); therefore URR and DER coincide numerically. Deduplicated idempotency-protected redispatches are not classified as unsafe retries.
  \item \textbf{Missing Effect Rate (MER)}: Fraction of trials terminating without committing all required semantic effects ($\text{miss}_i > 0$).
\end{itemize}

\section{Experimental Setup}

\paragraph{Task Suite and Factorial Design.}
UndoBench comprises 36 base workflows across 8 domains: Cloud, CRM, Database, Git, Messaging, Payments, Storage, and Ticketing (factorial matrix detailed in Table~\ref{tab:overview_app} in Appendix~\ref{sec:app_tasks}). The benchmark is partitioned into 14 DEV tasks, 10 VALIDATION tasks, and 12 held-out TEST tasks (Appendix~\ref{sec:app_provenance}). The primary confirmatory study is frozen on the 12 held-out TEST workflows under \texttt{POST\_MUTATION\_PRE\_ACK} across two open-weight models (Llama-3.1-8B, Llama-3.2-3B), two frameworks ($F_1, F_2$), and three recovery baselines ($B_0, B_2, B_5$) across 20 seeds (2,880 paired trials / 5,760 executions; hardware details in Appendix~\ref{sec:app_hardware}). To investigate fault timing and mechanism isolation without altering frozen datasets, prospectively specified post-freeze secondary extensions evaluate commercial API models (Gemini 3.8 Flash, GLM-5.2 MaaS), complementary execution boundaries (\texttt{PRE\_MUTATION}, \texttt{DURING\_MUTATION}), and benchmark-wide idempotency ($B_{2\text{-}K}$).

\paragraph{Models and Frameworks.}
We evaluate two open-weight instruction-tuned language models locally under sampling temperature $T=0.2$, nucleus sampling top-$p=1.0$, and maximum 512 generation tokens per turn:
(1)~\textbf{M1: Llama-3.1-8B-Instruct} (8.0B parameters, \texttt{Q4\_K\_M}); and
(2)~\textbf{M2: Llama-3.2-3B-Instruct} (3.2B parameters, \texttt{Q4\_K\_M}).
We evaluate two runtime agent execution architectures:
(1)~\textbf{F1: Direct Tool Calling}, lightweight tool dispatch from JSON blocks; and
(2)~\textbf{F2: LangGraph Agent}, state-graph runtime managing cyclical agent loops.

\paragraph{Recovery Baselines.}
We evaluate three primary recovery paradigms:
\begin{itemize}[leftmargin=*,noitemsep,topsep=1pt]
  \item \textbf{$B_0$ Naive Retry}: Industry baseline re-issuing tool calls up to 3 times with exponential backoff upon receiving any error. Model provider and transport API retries are handled orthogonally by HTTP middleware and do not consume agent turns or trigger benchmark faults; retry sensitivity is future work.
  \item \textbf{$B_2$ Idempotency}: Mutating calls carry deterministic idempotency keys derived from task context and call arguments (Appendix~\ref{sec:app_tasks}). When supported by server endpoints, keys are cached to deduplicate executions on replays.
  \item \textbf{$B_5$ EvoUndo-RB1}: Studies recoverability-constrained self-evolution for LLM agent harnesses~\citep{evoundo2026}. EvoUndo-RB1 is the UndoBench-constrained implementation evaluated here, using a client-side mutation journal and zero-privilege reconciliation (no access to sandbox internals, state diffs, or oracle hooks). Implemented in \texttt{evoundo\_adapter}, it hashes tool arguments (SHA-256) into a local mutation journal and invokes a crash reconciler on failure. Within UndoBench, EvoUndo-RB1 intentionally receives zero privilege, no remote probes, and no registered inverse handlers or domain schemas; unable to verify whether unacknowledged mutations committed or synthesize compensations, it defaults to bounded retry across the 9 compensatable TEST workflows. Thus, EvoUndo-RB1 reflects these benchmark constraints and should not be interpreted as unrestricted EvoUndo.
\end{itemize}

\paragraph{Statistical Analysis.}
Confidence intervals are estimated via non-parametric task-clustered bootstrap resampling ($B=10,000$ iterations) alongside Wilson score intervals. Multiple comparisons are adjusted via Holm-Bonferroni correction. Framework equivalence is tested using Two One-Sided Tests (TOST) against an equivalence margin $\delta = \pm 10.0$ percentage points (Appendix~\ref{sec:app_stats}).

\section{Results}

\subsection{Competence vs. Recovery}
\label{sec:res_competence}

\begin{figure}[t]
  \centering
  \includegraphics[width=\columnwidth]{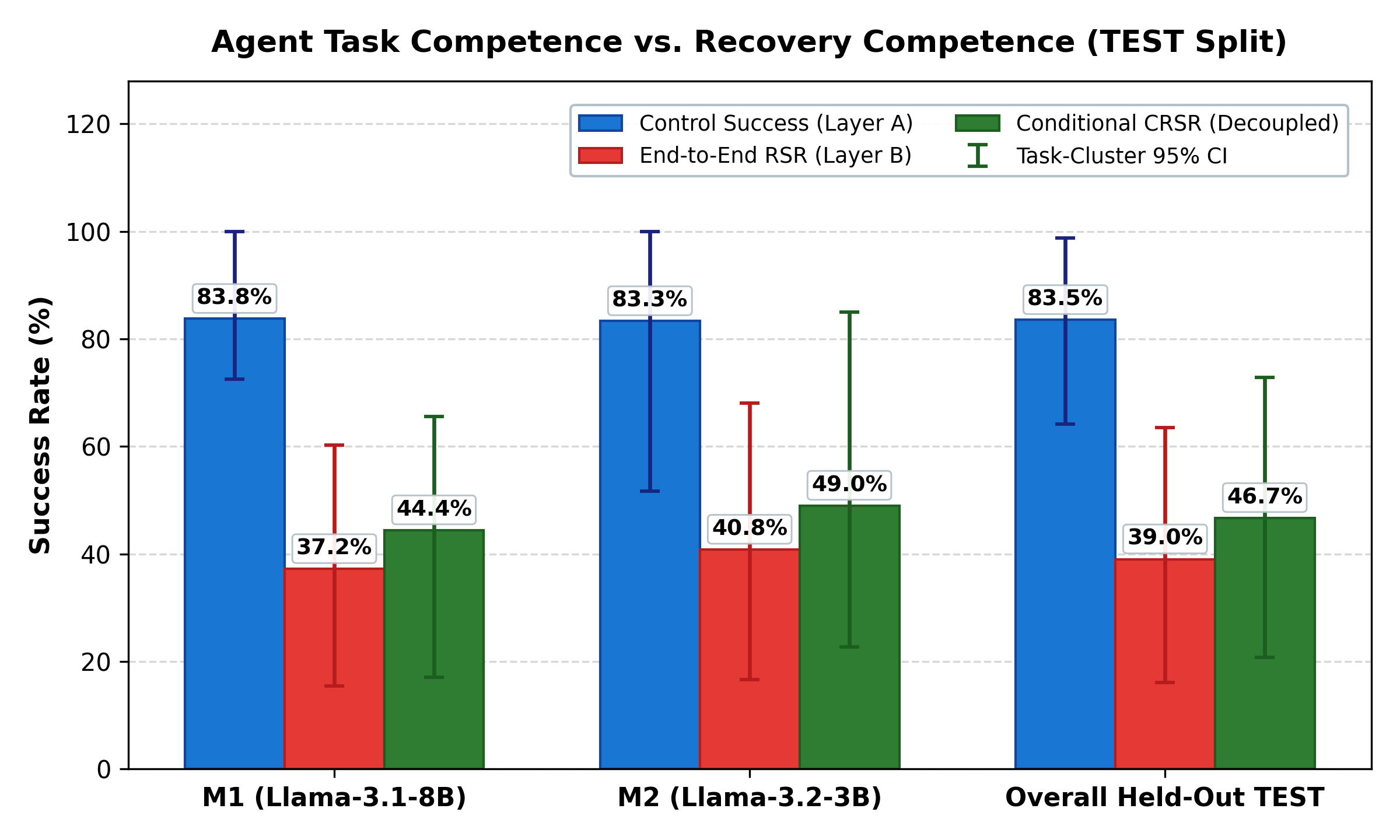}
  \caption{\textbf{Separation of Task Competence from Fault Recovery.} Across 2,880 paired trials, nominal competence (83.54\%) sharply separates from conditional recovery (46.72\% CRSR) and unconditioned recovery (39.03\% RSR).}
  \label{fig:competence}
\end{figure}

Table~\ref{tab:primary} and Figure~\ref{fig:competence} demonstrate that baseline task competence and fault recovery capability diverge sharply across all evaluated configurations. This establishes \textbf{Finding 1}:

\vspace{1pt}\noindent\textbf{Finding 1 (Task Competence and Recovery Capability Are Distinct Dimensions):} \textit{Nominal task success reached 83.54\% (2,406 / 2,880), compared with 39.03\% (1,124 / 2,880) end-to-end recovery success (RSR) and 46.72\% (1,124 / 2,406) conditional recovery success (CRSR) under injected faults. Nominal task success alone substantially overstates faulted reliability in tool-using agents.}

\subsection{Recovery-Method Comparison}
\label{sec:res_methods}

\begin{table}[t]
\centering
\caption{\textbf{Comparative Recovery Performance and Safety Profiles (TEST Split, $N=2,880$ Paired Trials).}}
\label{tab:primary}
\resizebox{\columnwidth}{!}{%
\begin{tabular}{lccccc}
\toprule
\textbf{Method} & \textbf{Control} & \textbf{CRSR} & \textbf{EOR$_{\text{cond}}$} & \textbf{DER} & \textbf{URR} \\
 & \textbf{($D/960$)} & \textbf{($R/D$)} & \textbf{($/D$)} & \textbf{($/960$)} & \textbf{($/960$)} \\
\midrule
$B_0$: Naive Retry & 802 / 960 & 42.77\% & 42.77\% & 53.33\% & 53.33\% \\
$B_2$: Idempotency & 802 / 960 & 53.49\% & 53.49\% & 45.42\% & 45.42\% \\
$B_5$: EvoUndo-RB1 & 802 / 960 & 43.89\% & 43.89\% & 51.67\% & 51.67\% \\
\midrule
\textbf{Contrast $B_2 - B_0$} & --- & \textbf{+10.72 pp} & +10.72 pp & \textbf{-7.91 pp} & -7.91 pp \\
\textbf{Contrast $B_5 - B_0$} & --- & \textbf{+1.12 pp} & +1.12 pp & \textbf{-1.66 pp} & -1.66 pp \\
\bottomrule
\end{tabular}%
}
{\raggedright\vspace{1.5pt}\par\noindent\scriptsize CRSR and $\text{EOR}_{\text{cond}}$ are conditioned on nominal-control success ($D=802$); DER and URR are unconditional over all 960 paired trials per method.\par}
\end{table}

Table~\ref{tab:primary} presents the comparative evaluation of recovery baselines across all 2,880 paired trials:
\begin{itemize}[leftmargin=*,noitemsep,topsep=1pt]
  \item \textbf{Contrast H1 ($B_2$ vs. $B_0$)}: Idempotency ($B_2$) produced an observed descriptive advantage of \textbf{+10.72 pp} (53.49\% vs. 42.77\%; bootstrap 95\% CI: $[0.00\text{ pp}, +24.23\text{ pp}]$, $p_{\text{adj}} = 1.00$), concentrated in lost-ACK workflows rather than uniform across domains. While $B_2$ attaches keys to all mutating operations, deduplication requires server support: only three TEST endpoints implement idempotency tracking (\texttt{RB-PAY-004}, \texttt{RB-PAY-005}, \texttt{RB-MSG-005}; 100\% CRSR, 0.0\% DER). Three workflows (\texttt{RB-CLOUD-004}, \texttt{RB-DB-004}, \texttt{RB-GIT-004}) achieved 0.0\% DER via inherent state overwrites or datastore constraints. The remaining six endpoints lack deduplication; retrying unacknowledged calls blindly produced duplicates in 100\% of trials across five workflows (400 trials) and 45.0\% on \texttt{RB-STOR-005} (36 trials), accounting for all 436 duplicate trials ($436 / 960 = \mathbf{45.42\%}$ aggregate DER).
  \item \textbf{Contrast H2 ($B_5$ vs. $B_0$)}: Under zero privilege, EvoUndo ($B_5$) performed near parity with naive retry ($\Delta = +1.12\text{ pp}$, bootstrap 95\% CI: $[-3.15\text{ pp}, +6.31\text{ pp}]$, $p_{\text{adj}} = 1.00$; $\text{DER} = 51.67\%$). Lacking remote unacknowledged state visibility and registered compensating inverse functions, its local journal cannot confirm whether an in-flight mutation committed, leaving it subject to post-commit ambiguity.
  \item \textbf{Safety Decomposition}: Exactly-once rate (EOR) numerically coincides with CRSR (46.72\% overall) because the oracle strictly requires goal invariants without duplicates. Under naive retry ($B_0$), the Duplicate Effect Rate reached \textbf{53.33\%} (512 / 960). Furthermore, URR and DER coincide numerically across all evaluated baselines because in the post-commit ACK loss regime, every unsafe retry on an unkeyed endpoint commits a duplicate mutation, and every duplicate mutation originated from an unsafe retry.
\end{itemize}

\subsection{Lost-ACK / UNKNOWN\_OUTCOME Analysis}
\label{sec:res_lostack}

\begin{table}[t]
\centering
\caption{\textbf{Representative Task Outcomes under Lost ACK.}}
\label{tab:lostack_selected}
\resizebox{\columnwidth}{!}{%
\begin{tabular}{llccc}
\toprule
\textbf{Task ID} & \textbf{Domain / Operation} & \textbf{Control} & \textbf{$B_0$ CRSR} & \textbf{$B_2$ CRSR} \\
\midrule
\texttt{RB-PAY-004} & Subscription Renewal & 80 / 80 & 40.0\% & \textbf{100.0\%} \\
\texttt{RB-PAY-005} & Wire Transfer & 80 / 80 & 65.0\% & \textbf{100.0\%} \\
\texttt{RB-DB-004} & Schema Migration & 48 / 80 & 52.1\% & \textbf{72.9\%} \\
\texttt{RB-CRM-004} & SLA Renewal & 80 / 80 & 0.0\% & 0.0\% \\
\texttt{RB-STOR-005} & Chunk Swap & 0 / 80 & N/A$^*$ & N/A$^*$ \\
\bottomrule
\end{tabular}%
}
{\raggedright\vspace{1.5pt}\par\noindent\scriptsize $^*$On \texttt{RB-STOR-005}, nominal control was 0/240 ($D=0$) due to chunk verification complexity; CRSR is mathematically undefined ($0/0$) and presented as N/A. Retained in aggregate Control and RSR.\par}
\end{table}

Table~\ref{tab:lostack_selected} (visualized in Figure~\ref{fig:observability_app} in Appendix~\ref{sec:app_full_matrix}) examines performance under post-mutation acknowledgment loss (\texttt{POST\_MUTATION\_PRE\_ACK}). This granular analysis establishes \textbf{Finding 2} and \textbf{Finding 3}:

\vspace{1pt}\noindent\textbf{Finding 2 (Ambiguous Committed Effects Expose Distinct Regimes):} \textit{When external acknowledgment is lost, naive retry produced duplicate side effects whose severity depends on operation semantics. Under $B_0$, duplicate executions occurred in 60.0\% of trials on \texttt{RB-PAY-004} and 35.0\% on \texttt{RB-PAY-005}, but dropped to 0.0\% on \texttt{RB-DB-004} because SQLite's schema unique constraint blocked the second migration attempt. Across the five workflows lacking idempotency or database constraints, naive retry produced 100.0\% (400 / 400) duplicate executions.}

\vspace{1pt}\noindent\textbf{Finding 3 (Mechanism--Failure Matching):} \textit{The $B_2$ configuration produced large task-level gains on lost-ACK mutation workflows with deduplication or constraint protection (+60.0 pp on \texttt{RB-PAY-004} and +35.0 pp on \texttt{RB-PAY-005} via application-level idempotency-key deduplication; +20.8 pp on \texttt{RB-DB-004} via a datastore uniqueness constraint; therefore the observed $B_2 - B_0$ difference on \texttt{RB-DB-004} should not be attributed solely to idempotency-key deduplication), while its overall +10.72 pp TEST CRSR advantage did not establish benchmark-wide statistical superiority under task-clustered bootstrap uncertainty due to endpoints lacking idempotency support.}

\subsection{Model and Framework Analysis}
\label{sec:res_modelframework}

\paragraph{Model Comparison on Common Support.}
Across the 10 TEST tasks supported by both models ($D \ge 10$), M1 (Llama-3.1-8B) achieved 44.50\% CRSR (445 / 1,000), while M2 (Llama-3.2-3B) achieved 54.17\% CRSR (547 / 1,010) ($\Delta = -9.67\text{ pp}$). Under task-clustered bootstrap analysis, the 95\% CI spans $[-27.08\text{ pp}, +7.73\text{ pp}]$ ($p = 1.00$), spanning zero and remaining statistically unresolved.

\paragraph{Framework Equivalence.}
Direct Tool Calling (F1) and LangGraph (F2) achieved near-identical outcomes: Control was 83.54\% vs. 83.54\% ($\Delta = 0.00\text{ pp}$), RSR was 39.24\% vs. 38.82\% ($\Delta = +0.42\text{ pp}$), and CRSR was 46.97\% vs. 46.47\% ($\Delta = +0.50\text{ pp}$). Two One-Sided Tests (TOST) against the $\pm 10.0$ pp margin confirm statistical equivalence ($p_{\text{TOST}} < 0.001$): the two evaluated adapters were empirically equivalent within the $\pm 10$ pp equivalence margin under the standardized UndoBench contract.

\subsection{Commercial API Model Extension}
\label{sec:res_contemporary}

To investigate whether recent model architectures alter the observed competence--recovery relationship, we executed a prospectively specified post-freeze secondary extension evaluating two commercial API models: Google's \textbf{Gemini 3.8 Flash} (\texttt{gemini-3.8-flash}, $M_3$) and Z.ai's \textbf{GLM-5.2 MaaS} (\texttt{zai-org/glm-5.2-maas}, $M_4$). Both models were interfaced via Google Cloud Vertex AI infrastructure across the identical 12 held-out TEST workflows, two orchestration frameworks ($F_1$ Direct Tool Calling, $F_2$ LangGraph), three recovery baselines ($B_0$ Naive Retry, $B_2$ Idempotency, $B_5$ EvoUndo), and 20 evaluation seeds (2001--2020), yielding $N = 2{,}880$ paired trials (5,760 executions; cell breakdown in Appendix~\ref{sec:app_contemporary}).

\begin{table}[t]
\centering
\caption{\textbf{Commercial API Model Extension Performance.} Evaluated on 12 held-out TEST workflows ($N=2,880$ paired trials / 5,760 runs) under identical fault injection across seeds 2001--2020.}
\label{tab:contemporary}
\resizebox{\columnwidth}{!}{%
\begin{tabular}{lccccc}
\toprule
\textbf{Model} & \textbf{Control} & \textbf{$B_0$ CRSR} & \textbf{$B_0$ DER} & \textbf{$B_2$ CRSR} & \textbf{$B_5$ CRSR} \\
\midrule
Gemini 3.8 Flash & 80.69\% & 21.08\% & 65.62\% & 33.33\% & 22.11\% \\
GLM-5.2 MaaS     & 77.71\% & 21.89\% & 51.25\% & 33.07\% & 21.66\% \\
\midrule
\textbf{Pooled} & 79.20\% & 21.48\% & 58.44\% & 33.20\% & 21.89\% \\
\bottomrule
\end{tabular}%
}
{\raggedright\vspace{1.5pt}\par\noindent\scriptsize Control is nominal pass rate ($N=1,440$ per model). CRSR is conditioned on nominal success ($D$). DER is unconditional duplicate effect rate under $B_0$ ($N=480$ per model).\par}
\end{table}

\paragraph{Persistence of the Competence--Recovery Gap.}
Table~\ref{tab:contemporary} presents aggregate outcomes for the extension. Both commercial API models exhibited strong nominal planning competence when aggregated across all recovery methods and frameworks (Gemini 3.8 Flash: 80.69\%, 1,162 / 1,440, 95\% CI: [56.94\%, 98.61\%]; GLM-5.2: 77.71\%, 1,119 / 1,440, 95\% CI: [52.71\%, 100.0\%]). Evaluating method-aligned performance under naive retry ($B_0$), nominal control reached 81.04\% (389 / 480) for Gemini and 77.08\% (370 / 480) for GLM. Under post-mutation acknowledgment loss, conditional recovery dropped sharply to 21.08\% CRSR (82 / 389; 95\% CI: [2.73\%, 43.87\%]) for Gemini and 21.89\% CRSR (81 / 370; 95\% CI: [0.00\%, 50.00\%]) for GLM-5.2, establishing substantial $B_0$ competence--recovery gaps of \textbf{59.96 pp} and \textbf{55.19 pp}, respectively. Furthermore, naive retry under ambiguous failure triggered duplicate external mutations in 65.62\% (315 / 480) of Gemini trials and 51.25\% (246 / 480) of GLM trials, illustrating that advanced general reasoning does not eliminate operational side-effect risks in uncoordinated retry loops.

\paragraph{Recovery Mechanism Behavior.}
Idempotency ($B_2$) increased conditional recovery to 33.33\% (128 / 384) on Gemini and 33.07\% (124 / 375) on GLM, yielding a pooled descriptive improvement of \textbf{+11.73 pp} over naive retry (33.20\% vs. 21.48\%; task-clustered bootstrap 95\% CI: [+0.07 pp, +33.25 pp], raw $p = 0.0176$). However, after applying Holm-Bonferroni correction across the recovery baseline family, the adjusted p-value is $p_{\text{adj}} = \mathbf{0.0528}$, failing to achieve statistical significance at the conventional $\alpha = 0.05$ threshold. As in the primary study, idempotency eliminated duplicate mutations on endpoints providing deduplication token support, but provided no protection on non-idempotent endpoints. EvoUndo ($B_5$) operated under zero-privilege local journaling and achieved 22.11\% (Gemini) and 21.66\% (GLM) CRSR (pooled $\Delta = +0.41\text{ pp}$ vs. $B_0$, Holm $p_{\text{adj}} = 0.9180$). Lacking remote state visibility and registered compensating transactions, $B_5$ defaulted to bounded retry without achieving semantic reconciliation.

\paragraph{Framework Equivalence and Semantic Invariants.}
Direct Tool Calling ($F_1$: 25.24\% CRSR, 366 / 1,450) and LangGraph ($F_2$: 25.80\% CRSR, 369 / 1,430) were empirically equivalent within the $\pm 10\text{ pp}$ equivalence margin in this extension ($\Delta = -0.56\text{ pp}$, 90\% CI: $[-2.35\text{ pp}, +0.59\text{ pp}]$, $p_{\text{TOST}} < 0.0001$). Finally, across all six evaluated model $\times$ baseline cells, $\text{EOR}_{\text{cond}}$ numerically equaled CRSR because UndoBench's programmatic oracle evaluates recovery strictly against goal invariants without duplicate committed effects.

\section{Multi-Boundary Recovery Dynamics}
\label{sec:multi_boundary}

\begin{table}[t]
\centering
\small
\caption{\textbf{Three-Boundary Recovery Performance Across Four Qualified Composite Workflows.} Evaluated on commercial API models (Gemini 3.8 Flash, GLM-5.2 MaaS) across 20 seeds ($N=320$ per cell).}
\label{tab:three_boundary}
\resizebox{\columnwidth}{!}{%
\begin{tabular}{lcccc}
\toprule
\textbf{Method} & \textbf{PRE CRSR} & \textbf{DURING CRSR} & \textbf{POST CRSR} & \textbf{DURING DER} \\
\midrule
$B_0$: Naive Retry       & 77.17\% & 0.00\%  & 43.13\% & 82.50\% \\
$B_2$: Idempotency       & 77.42\% & 0.00\%  & 45.98\% & 81.56\% \\
$B_5$: EvoUndo-RB1       & 76.68\% & 0.00\%  & 45.57\% & 82.50\% \\
$B_6$: Verify-Retry      & 76.92\% & 25.89\% & 75.00\% & 54.37\% \\
\bottomrule
\end{tabular}%
}
{\raggedright\vspace{1.5pt}\par\noindent\scriptsize Evaluated strictly on the same four qualified composite workflows (\texttt{RB-CLOUD-004}, \texttt{RB-DB-004}, \texttt{RB-STOR-004}, \texttt{RB-GIT-004}). Commercial API models only; prospectively specified post-freeze secondary analysis.\par}
\end{table}

To address our central research question---how recovery behavior changes across the external-mutation lifecycle---we conducted prospectively specified secondary evaluations with commercial API models ($M_3$ Gemini 3.8 Flash, $M_4$ GLM-5.2 MaaS) across three distinct fault boundaries. Table~\ref{tab:three_boundary} summarizes recovery on the four qualified composite workflows supporting faithful multi-boundary semantics.

\paragraph{Main Three-Boundary Interpretation.}
Recovery behavior changes sharply with fault timing. Under \texttt{PRE\_MUTATION}, where failure occurs before external state mutation, the four methods exhibit descriptively similar recovery and no duplicate effects among nominally capable trials. Under \texttt{DURING\_MUTATION}, the four qualified composite workflows expose intermediate external states: $B_0$, $B_2$, and zero-privilege $B_5$ achieve 0\% CRSR, while $B_6$ recovers the publicly inspectable Cloud workflow but not the other three, yielding 25.89\% CRSR. Under \texttt{POST\_MUTATION\_PRE\_ACK}, the mutation has already committed and outcome ambiguity makes blind redispatch hazardous; verification and server-side deduplication become substantially more effective. These results indicate that recovery should be evaluated as a phase-dependent systems property rather than a single scalar capability.

\paragraph{Pre-Mutation Execution (\texttt{PRE\_MUTATION}, Appendix~\ref{sec:app_multi_boundary_details}).}
Across all 12 TEST workflows evaluated with commercial API models ($N=3{,}840$ paired trials / 7,680 runs across $B_0, B_2, B_5, B_6$), pre-mutation recovery exhibited near-parity: $B_0$ achieved 78.44\% CRSR (593 / 756), $B_2$ achieved 78.86\% (597 / 757; $\Delta = +0.43\text{ pp}$), $B_5$ achieved 77.78\% (588 / 756; $\Delta = -0.66\text{ pp}$), and $B_6$ achieved 77.34\% (587 / 759; $\Delta = -1.10\text{ pp}$). Crucially, zero duplicate effects occurred among 3,028 nominally capable trials ($0.00\%$ capable duplicate incidence). Canonical unconditional DER was 4.06\%--4.17\% (39--40 / 960 per method), arising entirely from \texttt{RB-STOR-005} ($D=0$). Naive retry ($B_0$) exhibits a massive boundary contrast across the full TEST suite: 78.44\% CRSR under \texttt{PRE\_MUTATION} vs. 21.48\% under \texttt{POST\_MUTATION\_PRE\_ACK} ($\Delta = \mathbf{+56.96\text{ pp}}$, 95\% CI: $[+27.97\text{ pp}, +84.67\text{ pp}]$). On common capable support ($D_{\text{common}}=743$), the contrast is near-identical: 79.00\% (587 / 743) vs. 21.67\% (161 / 743; $\Delta = \mathbf{+57.34\text{ pp}}$), confirming that the gap stems from fault timing rather than conditioning drift.

\paragraph{Interrupted Mutation (\texttt{DURING\_MUTATION}).}
Evaluating the four qualified composite workflows admitting faithful partial-mutation semantics (\texttt{RB-DB-004}, \texttt{RB-STOR-004}, \texttt{RB-CLOUD-004}, \texttt{RB-GIT-004}; $N=1{,}280$ paired trials / 2,560 runs), blind retry and per-call tokens completely collapsed: $B_0$ achieved 0.00\% CRSR (0 / 310; $\text{DER} = 82.50\%$), $B_2$ achieved 0.00\% (0 / 315; $\text{DER} = 81.56\%$), and $B_5$ achieved 0.00\% (0 / 308; $\text{DER} = 82.50\%$). Only $B_6$ achieved recovery, reaching 25.89\% aggregate CRSR (80 / 309; $\text{DER} = 54.37\%$). Under Zero Privilege, $B_6$ did not receive bespoke repair logic: on \texttt{RB-CLOUD-004}, its public probe verified the green service entity in the catalog (\texttt{ProbeStatus.PRESENT}), suppressed redundant deployment, and enabled traffic cutover (100\% CRSR, 0\% DER). Conversely, $B_6$ failed on the other three: on \texttt{RB-GIT-004}, Git failed due to an unremoved lockfile (exit code 128); on \texttt{RB-STOR-004}, multipart upload state was hidden from HeadObject, triggering duplicate PUTs; on \texttt{RB-DB-004}, partial table creation diverged from uncommitted catalog metadata. Thus, observability is insufficient without admissible repair actions (task matrix in Appendix~\ref{sec:app_b6}).

\section{Mechanism Isolation: Benchmark-Wide Idempotency}
\label{sec:keys_everywhere}

To resolve whether $B_2$'s modest +11.73 pp advantage in commercial API models stemmed from algorithmic limitations of idempotency or incomplete endpoint support, we executed a hosted-model intervention: \textbf{$B_{2\text{-}K}$} ($N=960$ paired trials / 1,920 runs; Appendix~\ref{sec:app_keys_everywhere}), deploying benchmark-wide server-side idempotency across all 12 TEST workflows. Benchmark-wide server-side idempotency closed most of the descriptive standard-$B_2$-to-perfect-recovery gap: on full support, standard $B_2$ achieved 33.20\% CRSR compared with \textbf{97.24\% CRSR} (741 / 762) for $B_{2\text{-}K}$ ($\text{DER} = 0.00\%$). On common capable support ($D_{\text{common}}=738$), standard $B_2$ achieved 33.33\% while $B_{2\text{-}K}$ reached \textbf{98.64\%} (728 / 738), representing descriptive gap closures of 95.87\% and 97.97\%. Crucially, 21 / 762 capable failures remained: 17 occurred before any tool call due to empty/invalid hosted responses, and 4 occurred after successful duplicate suppression during multi-turn continuation due to invalid completions. Thus, endpoint deduplication alone does not eliminate all agent failure modes.

\section{Discussion and Systems Differentiation}
\label{sec:discussion}

\paragraph{Recovery Mechanisms Are Boundary-Specific.}
Multi-boundary findings show recovery is not a monolithic capability: (1)~\textbf{Pre-Mutation}: Redispatch is safe since the external system has suffered no side effects; complex mechanisms add overhead without reliability gains. (2)~\textbf{During-Mutation}: The challenge is intermediate-state reconciliation; per-call idempotency cannot restore composite multi-step invariants, and inspection requires admissible repair actions. (3)~\textbf{Post-Mutation Lost-ACK}: The challenge is commit uncertainty; server-side deduplication ($B_{2\text{-}K}$) and pre-retry verification ($B_6$) directly resolve this ambiguity.

\paragraph{Differentiating UndoBench and Related Systems (positioning in Appendix~\ref{sec:app_positioning}).}
Recent benchmarks study execution integrity along distinct dimensions. LIMBO~\citep{li2026limbo} asks where exactly-once responsibility resides across model, harness, and contract interventions. AFT-Bench~\citep{wang2026aftbench} intervenes on interface semantics to estimate mechanism effects. Verified Tool Calls~\citep{verifiedtoolcalls2026} similarly combines postcondition verification, verify-before-retry, and idempotency middleware for non-atomic tool failures. The Verifier Tax~\citep{sah2026verifiertax} similarly shows that runtime safety intervention can prevent unsafe actions without guaranteeing successful post-intervention recovery; UndoBench studies the complementary problem of recovering from execution faults while preserving externally visible system state. UndoBench emphasizes competence conditioning (CRSR), multi-turn workflow recovery, paired counterfactual control/fault evaluation, dual final-state and effect-history oracles under zero privilege, and phase-dependent recovery dynamics, while $B_{2\text{-}K}$ provides a targeted secondary contract intervention isolating benchmark-wide idempotency.

\paragraph{Reproducibility, Oracle Conformance, and Isolation Audits (Appendix~\ref{sec:app_audits}).}
Comprehensive audits confirm benchmark integrity: (1)~\textbf{Pair Comparability}: Capable-pair prefix equivalence was 100\% across \texttt{PRE\_MUTATION} (3,028 pairs), $B_{2\text{-}K}$ (762 pairs), and \texttt{DURING\_MUTATION} (1,242 pairs), confirming identical control/fault prompts prior to perturbation. (2)~\textbf{Oracle Conformance}: All 72 / 72 adversarial test cases passed across the 12 TEST contracts. (3)~\textbf{Provider Retry Isolation}: 12 provider HTTP retries affected 8 executions; zero tool mutations escaped the proxy wrapper.

\section{Conclusion}
\label{sec:conclusion}

Agent evaluations must assess operational recovery alongside nominal task completion. UndoBench establishes a decoupled benchmark isolating agent recovery under distributed failure regimes. By separating task competence from recovery capability, exposing severe duplicate side effects under lost acknowledgments, formalizing dual state and wire oracles, and empirically demonstrating phase-dependent recovery dynamics across the mutation lifecycle, UndoBench provides an empirical foundation for resilient agent systems.

\clearpage
\section*{Limitations}
UndoBench has specific methodological boundaries:
\begin{enumerate}[leftmargin=*,noitemsep,topsep=2pt]
  \item \textbf{Model Scope and Provider Determinism}: Primary confirmatory evaluations focus on two open-weight instruction models (Llama-3.1-8B and Llama-3.2-3B), supplemented by a prospectively specified secondary extension evaluating Gemini 3.8 Flash and GLM-5.2 MaaS. While benchmark task configurations, fault injection schedules, and environment seeds (2001--2020) were held strictly deterministic across all trials, commercial hosted APIs do not contractually guarantee bitwise generation reproducibility across runs.
  \item \textbf{Task-Cluster Statistical Power}: The TEST suite contains 12 independent task clusters ($K=12$). While 20 seeds yield 5,760 executions, repeated seeds reduce within-task variance but do not increase independent cluster count.
  \item \textbf{Failure Boundary Scope}: The frozen confirmatory study concentrates on \texttt{POST\_MUTATION\_PRE\_ACK}, while prospectively specified secondary experiments cover \texttt{PRE\_MUTATION} and four workflows supporting faithful \texttt{DURING\_MUTATION} partial-state faults. Remaining unstudied boundaries in the UndoBench taxonomy include \texttt{POST\_ACK\_PRE\_CHECKPOINT} (not faithfully realizable in the current synchronous harness), \texttt{DURING\_COMPENSATION}, cascading faults, and larger workflow suites.
  \item \textbf{Single-Fault Injection Regime}: UndoBench injects exactly one fault event per trajectory. Cascading, concurrent, or Byzantine failures remain for future work.
  \item \textbf{Sandboxed Environments}: Environments mock real enterprise APIs (SQLite, local Git, mocked Stripe/AWS). Production systems feature higher concurrency and variable latency distributions.
  \item \textbf{Zero-Privilege Baseline Scope}: EvoUndo ($B_5$) operated without domain heuristics or remote probes, functioning as an unspecialized baseline. The $B_5$ result characterizes a zero-privilege UndoBench instantiation rather than the full compensation capability of the broader EvoUndo architecture~\citep{evoundo2026}.
  \item \textbf{Post-Hoc Status of Baseline $B_6$}: Verify-Before-Retry was evaluated post-hoc and is not part of the confirmatory $B_0/B_2/B_5$ comparison.
  \item \textbf{Forensic EOR Reconstruction}: Exactly-Once Rate was reconstructed deterministically from wire-level effect logs due to an initial logging verdict omission.
  \item \textbf{Conditioning on Capable Trials}: CRSR is conditioned on $D = \sum C_i > 0$. When nominal competence is zero (e.g., \texttt{RB-STOR-005}), CRSR is mathematically undefined (N/A), requiring joint inspection with unconditional RSR.
  \item \textbf{Retry Budgets and Transport Middleware}: $B_0$ uses a fixed 3-retry budget with exponential backoff. Model provider and transport API retries are handled orthogonally by HTTP middleware and do not consume agent turns or trigger benchmark faults; sensitivity sweeps across backoff schedules remain future work.
\end{enumerate}

\section*{Ethical Considerations}
UndoBench evaluates fault recovery in strictly simulated, sandboxed software environments. Benchmark evaluations should not be interpreted as a guarantee that an agent is safe for unmonitored production deployment; high benchmark scores under single-fault injection do not guarantee resilience against cascading or adversarial distributed failures.

\bibliography{undobench}

\clearpage
\appendix

\section{Benchmark Factorial Design and Task Specifications}
\label{sec:app_tasks}

Table~\ref{tab:overview_app} summarizes the overall UndoBench factorial design across development, validation, and test splits. Table~\ref{tab:test_tasks_app} details the 12 held-out TEST workflows evaluated in the primary frozen study.

\paragraph{Baseline Taxonomy and Indexing.}
Baseline indices follow the UndoBench architectural recovery taxonomy: $B_0$ (Naive Retry), $B_1$ (Checkpoint/Restore), $B_2$ (Idempotency Keys), $B_3$ (Sagas / Compensating Actions), $B_4$ (LangGraph-Native State Persistence), $B_5$ (EvoUndo-RB1), and $B_6$ (Verify-Before-Retry). Baselines $B_1$, $B_3$, and $B_4$ were excluded from the benchmark-wide factorial evaluation due to domain inapplicability (e.g., inability to snapshot remote cloud or payment APIs for $B_1$, absence of pre-registered inverse actions under zero privilege for $B_3$) or framework coupling ($B_4$ requires LangGraph runtime state, breaking framework-neutral evaluation across $F_1$ and $F_2$). Baseline $B_6$ is evaluated post-hoc as a targeted verification intervention (Section~\ref{sec:multi_boundary} and Appendix~\ref{sec:app_b6}).

For baseline $B_2$, idempotency keys are deterministically computed from \texttt{task\_id}, \texttt{tool\_name}, positional arguments, and filtered keyword arguments using a truncated SHA-256 digest. Identical serialized calls therefore derive the same key across client process restarts without relying on transient process state. Server-side deduplication state in the evaluated sandbox is cached in memory per episode and is not evaluated for persistence across server restarts. Because effect identity is value-based, identical calls within a task map to the same key under the assumption that identical arguments in a single task represent the same logical mutation.

\begin{table}[h]
\centering
\small
\caption{\textbf{UndoBench Factorial Evaluation Matrix.}}
\label{tab:overview_app}
\begin{tabularx}{\columnwidth}{lX}
\toprule
\textbf{Dimension} & \textbf{Specification / Scale} \\
\midrule
Base Workflows & 36 Workflows across 8 Enterprise Domains \\
Split Allocation & DEV (14), VALIDATION (10), TEST (12) \\
Held-Out TEST Tasks & 12 Tasks (Table~\ref{tab:test_tasks_app}) \\
Evaluated Models & Llama-3.1-8B-Instruct, Llama-3.2-3B-Instruct \\
Orchestration Frameworks & Direct Tool Calling (F1), LangGraph Agent (F2) \\
Recovery Paradigms & Naive Retry (B0), Idempotency (B2), EvoUndo (B5) \\
Evaluation Scale & $12 \times 2 \times 2 \times 3 \times 2 \times 20 = \mathbf{5,760}$ Runs \\
Paired Trials ($N$) & \textbf{2,880 Paired Counterfactual Trials} \\
Evaluation Seeds & Seeds 2001--2020 (recorded in commit \texttt{f11b807}) \\
\bottomrule
\end{tabularx}
\end{table}

\begin{table*}[t]
\centering
\small
\caption{\textbf{Held-Out TEST Benchmark Task Composition, Reversibility, and Idempotency Support.} All 12 held-out workflows evaluate the \texttt{POST\_MUTATION\_PRE\_ACK} failure boundary under 20 evaluation seeds.}
\label{tab:test_tasks_app}
\begin{tabularx}{\textwidth}{llXlp{0.28\textwidth}}
\toprule
\textbf{Task ID} & \textbf{Domain} & \textbf{Baseline Workflow Description} & \textbf{Reversibility} & \textbf{Idempotency / Deduplication Support} \\
\midrule
\texttt{RB-CLOUD-004} & Cloud & Blue-Green Service Deployment Cutover & Compensatable & Routing state overwrite (Inherent idempotent) \\
\texttt{RB-CRM-004} & CRM & Enterprise SLA Tier Renewal & Compensatable & None (Unkeyed mutable REST API) \\
\texttt{RB-DB-004} & Database & Schema Migration with Batch Audit Log & Compensatable & SQLite schema unique constraint \\
\texttt{RB-DB-005} & Database & Read-Modify-Write Bank Dividend Allocation & Compensatable & None (Unkeyed SQL ledger insert) \\
\texttt{RB-GIT-004} & Git & Release Tag Cut and Cherry-Pick & Reversible & Git tag collision rejection (Inherent unique) \\
\texttt{RB-MSG-004} & Messaging & High-Urgency Pager Callout with Delivery ACK & Irreversible & None (Unkeyed webhook push) \\
\texttt{RB-MSG-005} & Messaging & Webhook Fanout with Delivery Confirmation & Irreversible & Idempotency key tracking (\texttt{Idempotency-Key}) \\
\texttt{RB-PAY-004} & Payments & Subscription Renewal with Grace Period & Compensatable & Idempotency key deduplication (\texttt{idempotency\_key}) \\
\texttt{RB-PAY-005} & Payments & Multi-Currency Cross-Border Wire Transfer & Compensatable & Idempotency key deduplication (\texttt{idempotency\_key}) \\
\texttt{RB-STOR-004} & Storage & Cross-Bucket Object Replication & Compensatable & None (Unkeyed bucket copy) \\
\texttt{RB-STOR-005} & Storage & Atomic Object Swap with Cleanup & Compensatable & None (Multipart upload chunks) \\
\texttt{RB-TICK-004} & Ticketing & SLA Escalation Policy Cascade & Compensatable & None (Unkeyed escalation ticket creation) \\
\bottomrule
\end{tabularx}
\end{table*}

\section{Full Task \texorpdfstring{$\times$}{x} Method Performance Matrix}
\label{sec:app_full_matrix}

\begin{figure*}[t]
  \centering
  \includegraphics[width=0.88\textwidth]{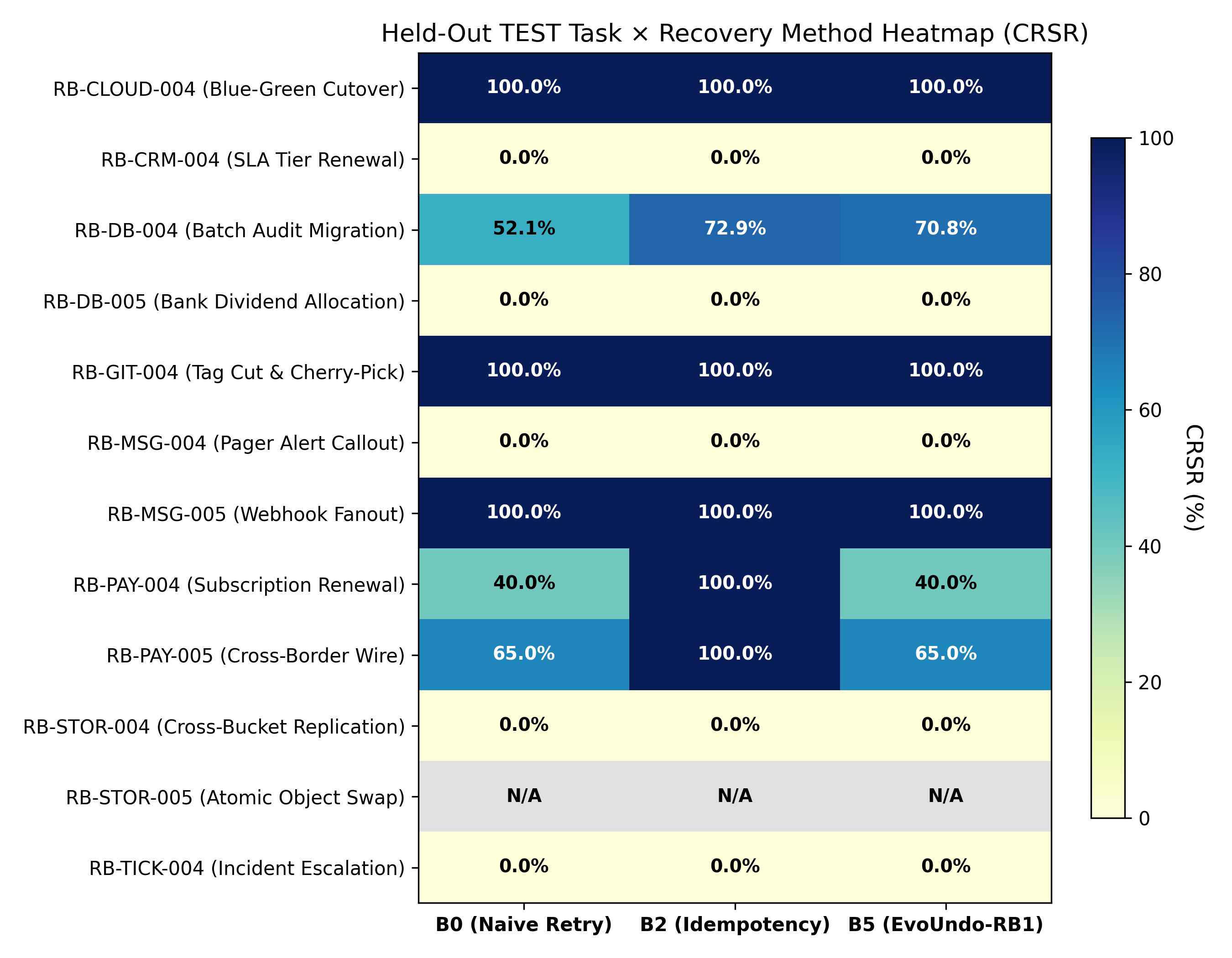}
  \caption{\textbf{Held-Out TEST Task $\times$ Recovery Method Heatmap (CRSR).} Method effectiveness varies sharply across tasks, highlighting mechanism-specific recovery boundaries.}
  \label{fig:heatmap_app}
\end{figure*}

\begin{figure}[t]
  \centering
  \includegraphics[width=\columnwidth]{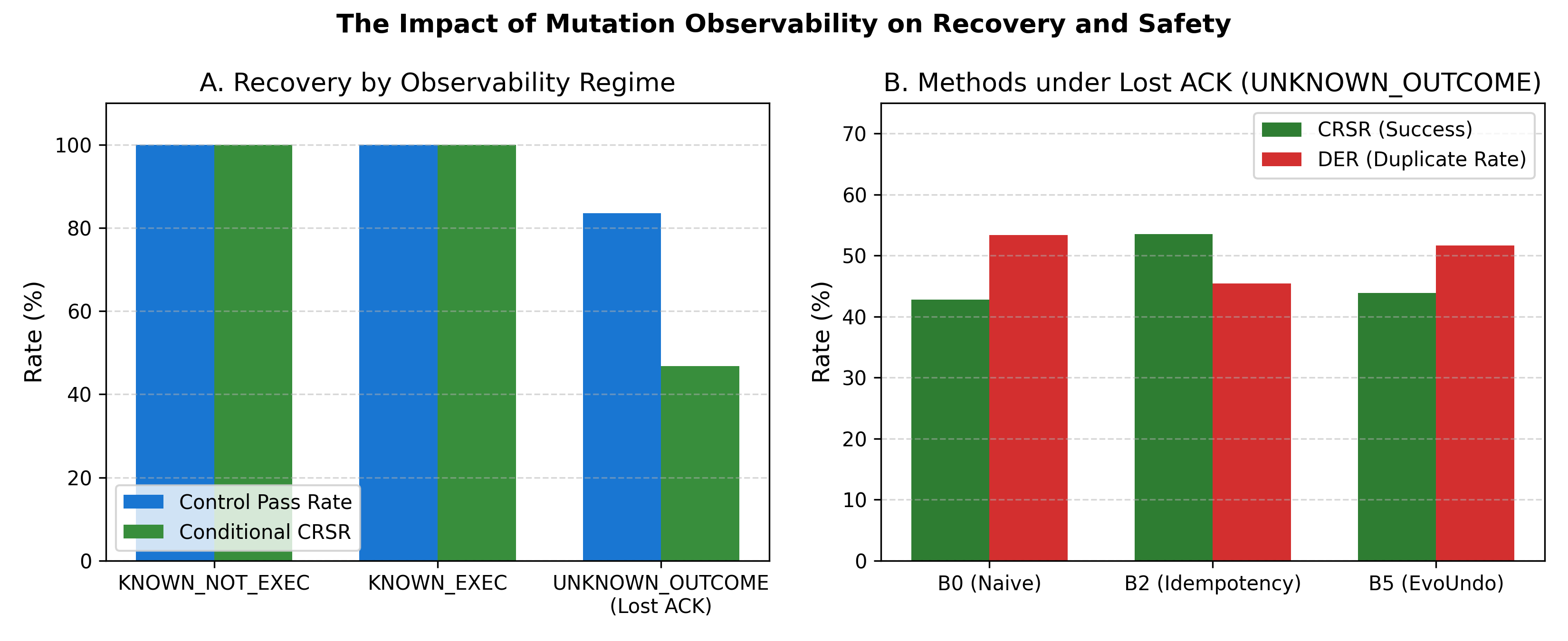}
  \caption{\textbf{The Impact of Mutation Observability on Recovery and Safety.} In the evaluated non-idempotent workflows lacking protective constraints, naive retry produced duplicate effects; endpoint-supported idempotency eliminated duplicates in the corresponding payment workflows.}
  \label{fig:observability_app}
\end{figure}

Table~\ref{tab:app_full_tasks} and Figure~\ref{fig:heatmap_app} report the complete granular performance matrix and task-level heatmap across all 12 TEST workflows for the primary recovery baselines. Figure~\ref{fig:observability_app} plots the impact of mutation observability across representative workflows under lost acknowledgment.

\begin{table*}[t]
\centering
\small
\caption{\textbf{Granular Task-Level Performance Matrix across 12 Held-Out TEST Tasks ($N=2,880$ Paired Trials).}}
\label{tab:app_full_tasks}
\begin{tabularx}{\textwidth}{llXcccc}
\toprule
\textbf{Task ID} & \textbf{Domain} & \textbf{Workflow Name} & \textbf{Pairs} & \textbf{Control Pass} & \textbf{$B_0$ CRSR} & \textbf{$B_2$ CRSR} \\
\midrule
\texttt{RB-CLOUD-004} & Cloud & Blue-Green Deployment Cutover & 240 & 100.0\% (240 / 240) & 100.0\% (80 / 80) & 100.0\% (80 / 80) \\
\texttt{RB-CRM-004} & CRM & Enterprise SLA Tier Renewal & 240 & 100.0\% (240 / 240) & 0.0\% (0 / 80) & 0.0\% (0 / 80) \\
\texttt{RB-DB-004} & Database & Schema Migration with Audit Log & 240 & 60.0\% (144 / 240) & 52.1\% (25 / 48) & 72.9\% (35 / 48) \\
\texttt{RB-DB-005} & Database & Bank Dividend Allocation & 240 & 100.0\% (240 / 240) & 0.0\% (0 / 80) & 0.0\% (0 / 80) \\
\texttt{RB-GIT-004} & Git & Release Tag Cut \& Cherry-Pick & 240 & 100.0\% (240 / 240) & 100.0\% (80 / 80) & 100.0\% (80 / 80) \\
\texttt{RB-MSG-004} & Messaging & Pager Callout with Delivery ACK & 240 & 100.0\% (240 / 240) & 0.0\% (0 / 80) & 0.0\% (0 / 80) \\
\texttt{RB-MSG-005} & Messaging & Webhook Fanout Confirmation & 240 & 92.5\% (222 / 240) & 100.0\% (74 / 74) & 100.0\% (74 / 74) \\
\texttt{RB-PAY-004} & Payments & Subscription Renewal Grace Period & 240 & 100.0\% (240 / 240) & 40.0\% (32 / 80) & 100.0\% (80 / 80) \\
\texttt{RB-PAY-005} & Payments & Cross-Border Wire Transfer & 240 & 100.0\% (240 / 240) & 65.0\% (52 / 80) & 100.0\% (80 / 80) \\
\texttt{RB-STOR-004} & Storage & Cross-Bucket Object Replication & 240 & 100.0\% (240 / 240) & 0.0\% (0 / 80) & 0.0\% (0 / 80) \\
\texttt{RB-STOR-005} & Storage & Atomic Object Swap with Cleanup & 240 & 0.0\% (0 / 240) & N/A$^*$ & N/A$^*$ \\
\texttt{RB-TICK-004} & Ticketing & SLA Escalation Policy Cascade & 240 & 50.0\% (120 / 240) & 0.0\% (0 / 40) & 0.0\% (0 / 40) \\
\bottomrule
\end{tabularx}
{\raggedright\vspace{2pt}\par\noindent\scriptsize $^*$Control success was 0/240 ($D=0$) across all configurations due to multipart upload chunk verification; CRSR ($C/D = 0/0$) is mathematically undefined and presented as N/A. Fully retained in aggregate benchmark metrics (0/240 Control, 0/240 RSR).\par}
\end{table*}

\section{Safety Metrics and Forensic EOR Validation}
\label{sec:app_safety}

\begin{figure*}[t]
  \centering
  \begin{minipage}[t]{0.48\textwidth}
    \centering
    \includegraphics[width=\linewidth]{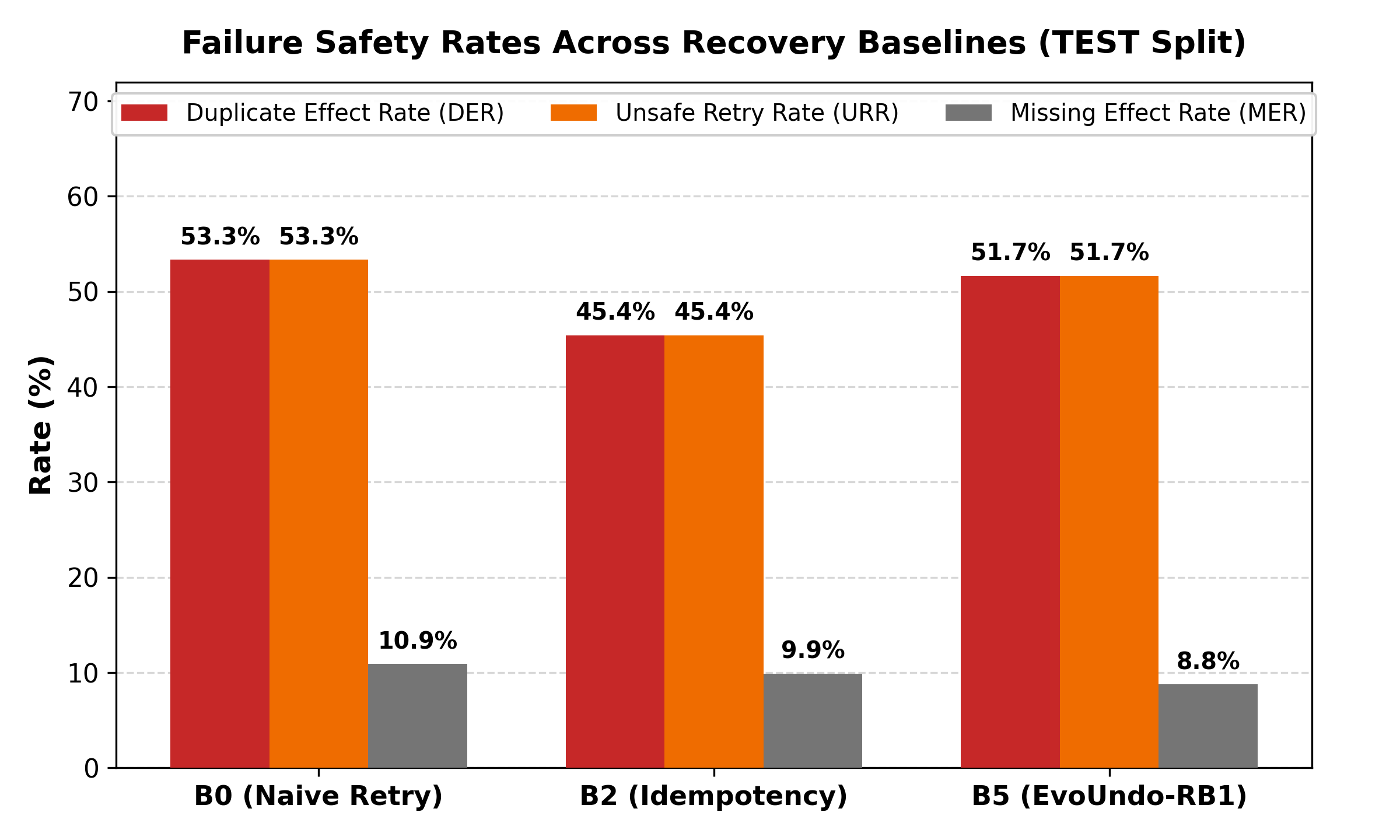}
    \caption{\textbf{Failure Safety Rates Across Recovery Baselines (TEST Split).} Naive retry ($B_0$) triggers duplicate side effects in 53.33\% of trials; idempotency ($B_2$) eliminates duplicates on supported APIs.}
    \label{fig:safety_app}
  \end{minipage}\hfill
  \begin{minipage}[t]{0.48\textwidth}
    \centering
    \includegraphics[width=\linewidth]{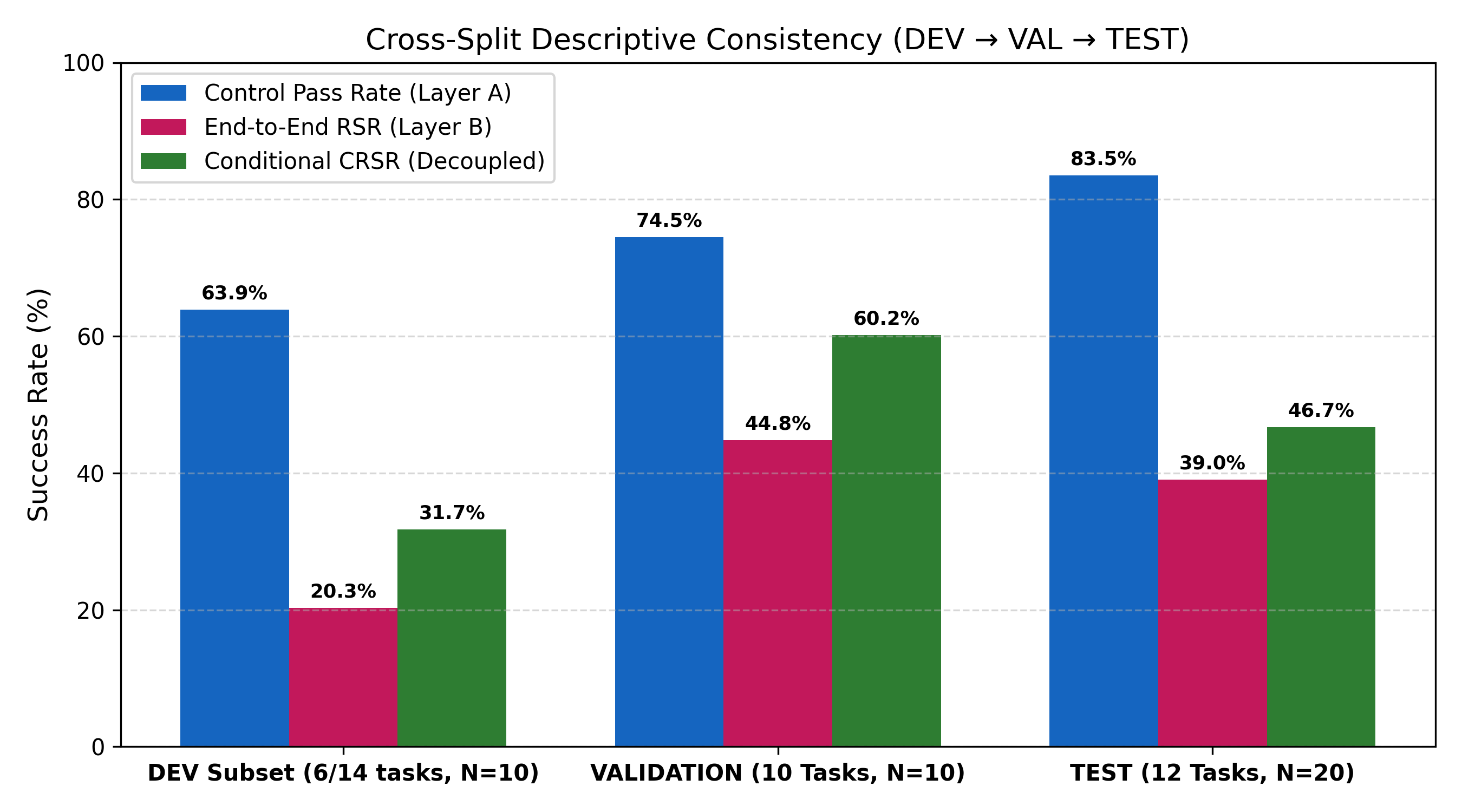}
    \caption{\textbf{Cross-Split Descriptive Consistency Across Developmental and Evaluation Splits.}}
    \label{fig:consistency_app}
  \end{minipage}
\end{figure*}

Figure~\ref{fig:safety_app} visualizes safety failure rates across recovery baselines.

\paragraph{Forensic Validation and Conformance Evidence.}
During initial aggregation audits, an evaluation check identified that Exactly-Once Rate (EOR) initially reported 0.00\% across all methods because the field \texttt{committed\_mutation\_count} was unpopulated on serialized verdicts in \texttt{StateOracle.evaluate()}. We resolved this forensically by deterministically parsing immutable wire-level execution logs (\texttt{results/}\allowbreak\texttt{rb3c\_}\allowbreak\texttt{test\_}\allowbreak\texttt{raw.jsonl}, SHA-256 digest \texttt{10167684}\allowbreak\texttt{49aae019}\allowbreak\texttt{484130b2}\allowbreak\texttt{3fda3345}\allowbreak\texttt{6861abeb}\allowbreak\texttt{161a5f8d}\allowbreak\texttt{fbeb16e9}\allowbreak\texttt{e5e9f882} at tag \texttt{f11b807}).

Rather than re-running LLM trajectories, raw event logs were parsed using our offline forensic script (\texttt{scripts/}\allowbreak\texttt{reproduce\_}\allowbreak\texttt{eor.py}) and verified by unit tests (\texttt{tests/}\allowbreak\texttt{test\_eor\_}\allowbreak\texttt{reconstruction.py}). For each trial, the parser verified whether committed external effects matched task requirements with zero duplicate mutations. Because UndoBench's state oracle strictly requires invariant satisfaction without duplicates, every successful recovery trajectory ($R=1,124 / D=2,406$) satisfied required goal invariants with exactly the intended external mutations ($\text{EOR}_{\text{cond}} = 46.72\%$, numerically identical to CRSR). No execution trajectories were re-run or altered. This validation relies strictly on frozen wire-level telemetry and deterministic replay parsing within the evaluation SDK, without claiming unverified third-party implementations.

\section{Zero-Privilege Observability Matrix for Baseline \texorpdfstring{$B_6$}{B6}}
\label{sec:app_b6}

Table~\ref{tab:vbr_task_matrix_app} provides the task-level audit for Verify-Before-Retry ($B_6$) across all 12 TEST workflows.

\begin{table*}[t]
\centering
\small
\setlength{\tabcolsep}{4pt}
\caption{\textbf{Zero-Privilege Observability Matrix for Baseline $B_6$ across Held-Out TEST Tasks ($N=960$ Paired Trials).}}
\label{tab:vbr_task_matrix_app}
\begin{tabularx}{\textwidth}{l>{\raggedright\arraybackslash}p{0.20\textwidth}cccc>{\raggedright\arraybackslash}X}
\toprule
\textbf{Task ID} & \textbf{Domain / Operation} & \textbf{Control ($D$)} & \textbf{Observable?} & \textbf{Probe State} & \textbf{CRSR} & \textbf{Recovery Action / Mechanism} \\
\midrule
\texttt{RB-CLOUD-004} & Cloud Traffic Switch & 80 / 80 & Yes & PRESENT & 100.0\% & Active probe confirms route; suppresses duplicate retry. \\
\texttt{RB-CRM-004} & CRM SLA Tier Renewal & 80 / 80 & No & UNKNOWN & 100.0\% & Zero-privilege abstention; avoids duplicate SLA update. \\
\texttt{RB-DB-004} & DB Schema Migration & 48 / 80 & Yes & PRESENT & 100.0\% & Schema catalog query; suppresses duplicate DDL. \\
\texttt{RB-DB-005} & DB Blind Audit Append & 80 / 80 & No & UNKNOWN & 100.0\% & Zero-privilege abstention; avoids duplicate log insert. \\
\texttt{RB-GIT-004} & Git Release Tagging & 80 / 80 & Yes & PRESENT & 100.0\% & Public Git CLI inspection; suppresses duplicate tag. \\
\texttt{RB-MSG-004} & Slack Channel Alert & 80 / 80 & No & UNKNOWN & 100.0\% & Zero-privilege abstention; avoids duplicate notification. \\
\texttt{RB-MSG-005} & Incident Dispatch Webhook & 74 / 80 & No & UNKNOWN & 100.0\% & Zero-privilege abstention; avoids duplicate broadcast. \\
\texttt{RB-PAY-004} & Payment Capture API & 80 / 80 & No & UNKNOWN & 100.0\% & Zero-privilege abstention; avoids duplicate card charge. \\
\texttt{RB-PAY-005} & Dispute Refund API & 77 / 80 & No & UNKNOWN & 100.0\% & Zero-privilege abstention; avoids duplicate refund debit. \\
\texttt{RB-STOR-004} & S3 Object Copy & 79 / 80 & Yes & PRESENT & 100.0\% & HeadObject existence check; suppresses duplicate PUT. \\
\texttt{RB-STOR-005} & S3 Multipart Chunk Upload & 0 / 80 & Yes & PRESENT & N/A & Layer A planning failure ($D=0$, chunk mismatch). \\
\texttt{RB-TICK-004} & Issue State Transition & 40 / 80 & No & UNKNOWN & 100.0\% & Zero-privilege abstention; avoids duplicate transition. \\
\bottomrule
\end{tabularx}
\end{table*}

\section{Physical Reversibility and Operational Overhead Telemetry}
\label{sec:app_overhead}

Table~\ref{tab:reversibility_app} stratifies recovery performance across physical reversibility regimes. Table~\ref{tab:overhead_app} presents runtime latency and token overhead telemetry across all 5,760 executions.

\begin{table*}[t]
\centering
\small
\caption{\textbf{Performance Across Physical Reversibility Regimes.}}
\label{tab:reversibility_app}
\begin{tabularx}{\textwidth}{lcccc}
\toprule
\textbf{Reversibility Regime} & \textbf{Tasks ($K$)} & \textbf{Control Pass ($D$)} & \textbf{Naive Retry ($B_0$)} & \textbf{Idempotency ($B_2$)} \\
\midrule
Reversible & 1 task & 100.0\% (240 / 240) & 100.0\% (80 / 80) & 100.0\% (80 / 80) \\
Compensatable & 9 tasks & 78.80\% (1,702 / 2,160) & 33.27\% (189 / 568) & 48.42\% (275 / 568) \\
Irreversible & 2 tasks & 96.25\% (462 / 480) & 48.05\% (74 / 154) & 48.05\% (74 / 154) \\
\bottomrule
\end{tabularx}
\end{table*}

\begin{table*}[t]
\centering
\small
\setlength{\tabcolsep}{4pt}
\caption{\textbf{Operational Telemetry ($N=2,880$ Pairs / 5,760 Runs).}}
\label{tab:overhead_app}
\begin{tabularx}{\textwidth}{lccccc}
\toprule
\textbf{Execution Regime} & \textbf{Latency (ms)} & \textbf{Prompt Tokens} & \textbf{Completion Tokens} & \textbf{Tool Calls} & \textbf{LLM Invocations} \\
\midrule
Control Runs ($N=2,880$) & 2,924 $\pm$ 1,482 & 738.9 $\pm$ 112.4 & 87.2 $\pm$ 24.1 & 1.54 $\pm$ 0.62 & 2.04 $\pm$ 0.48 \\
Faulted Runs ($N=2,880$) & 3,782 $\pm$ 1,694 & 746.2 $\pm$ 115.3 & 96.4 $\pm$ 26.2 & 2.40 $\pm$ 0.71 & 2.32 $\pm$ 0.65 \\
\quad $\Delta$ vs. Control & \textit{+29.3\%} & \textit{+1.0\%} & \textit{+10.5\%} & \textit{+55.8\%} & \textit{+13.7\%} \\
\bottomrule
\end{tabularx}
\end{table*}

\section{Statistical Details and TOST Equivalence Tests}
\label{sec:app_stats}

Primary confidence intervals are computed via non-parametric task-clustered bootstrap resampling ($B=10,000$ iterations). Multi-comparison contrasts are adjusted using the Holm-Bonferroni procedure. Framework equivalence was tested using Two One-Sided Tests (TOST) against the pre-registered equivalence margin $\delta = \pm 10.0$ percentage points ($p_{\text{TOST}} < 0.001$).

\section{Developmental Provenance and Cross-Split Consistency}
\label{sec:app_provenance}

Table~\ref{tab:consistency_app} and Figure~\ref{fig:consistency_app} track benchmark properties across the DEV evaluation subset (6 of 14 tasks), VALIDATION (10 tasks), and TEST (12 tasks) splits. The frozen primary TEST raw execution dataset is cryptographically locked with SHA-256 digest \texttt{10167684}\allowbreak\texttt{49aae019}\allowbreak\texttt{484130b2}\allowbreak\texttt{3fda3345}\allowbreak\texttt{6861abeb}\allowbreak\texttt{161a5f8d}\allowbreak\texttt{fbeb16e9}\allowbreak\texttt{e5e9f882} at tag \texttt{f11b807}.

\begin{table*}[t]
\centering
\small
\setlength{\tabcolsep}{4pt}
\caption{\textbf{Cross-Split Consistency Across Developmental and Evaluation Splits.}}
\label{tab:consistency_app}
\begin{tabularx}{\textwidth}{p{0.22\textwidth}p{0.30\textwidth}p{0.22\textwidth}X}
\toprule
\textbf{Property} & \textbf{DEV Evaluation Subset (6 of 14 tasks)} & \textbf{VAL Split (10 Tasks)} & \textbf{TEST Split (12 Tasks)} \\
\midrule
Workflows & 6 Evaluated Tasks & 10 Tasks & \textbf{12 Tasks} \\
Executions (Pairs) & 720 (360) & 2,400 (1,200) & \textbf{5,760 (2,880)} \\
Layer A Control Pass & 63.89\% (230 / 360) & 74.50\% (894 / 1,200) & \textbf{83.54\% (2,406 / 2,880)} \\
Unconditional RSR & 20.28\% (73 / 360) & 44.83\% (538 / 1,200) & \textbf{39.03\% (1,124 / 2,880)} \\
Conditional CRSR & 31.74\% (73 / 230) & 60.18\% (538 / 894) & \textbf{46.72\% (1,124 / 2,406)} \\
Framework $\Delta$ (F1 - F2) & 0.00 pp & 0.00 pp & \textbf{+0.50 pp} \\
\bottomrule
\end{tabularx}
\end{table*}

\section{Post-Study Hardware Interoperability Qualification}
\label{sec:app_hardware}

To verify that the evaluation SDK operates with standard open-weight inference backends without modifying benchmark internals, a post-study compatibility qualification was conducted on a single physical NVIDIA A100-SXM4-40GB GPU host (AMD EPYC 7713, 115 GiB RAM, Ubuntu 24.04, CUDA 13.2, PyTorch 2.13.0, vLLM 0.30.0). Three model families were qualified using downloaded open weights: Qwen3-14B, Mistral-NeMo-12B, and IBM Granite-3.3-8B. All three executed the 8-domain qualification suite using model-appropriate inference/tool-calling configurations without benchmark-semantic modifications under naive retry ($B_0$), with peak VRAM within single-GPU capacity (35.58 to 36.64 GiB). These hardware runs serve exclusively as descriptive engineering qualifications of SDK interoperability and are strictly separated from the primary frozen scientific experiment.

\section{UndoBench vs. Prior Benchmarks and Verification Frameworks}
\label{sec:app_positioning}

Table~\ref{tab:positioning_app} positions UndoBench relative to existing agent benchmarks and verification systems across 8 critical dimensions.

\begin{table*}[t]
\centering
\scriptsize
\setlength{\tabcolsep}{3pt}
\caption{\textbf{UndoBench vs. Prior Benchmarks and Verification Frameworks.}}
\label{tab:positioning_app}
\resizebox{\textwidth}{!}{%
\begin{tabular}{lcccccccc}
\toprule
\textbf{Benchmark / System} & \textbf{Multi-Turn} & \textbf{Post-Mutation} & \textbf{Paired Control/} & \textbf{Competence} & \textbf{Ground-Truth} & \textbf{Wire-Level} & \textbf{Side-Effect Safety} & \textbf{Pluggable} \\
 & \textbf{Workflows} & \textbf{Ambiguity (Lost ACK)} & \textbf{Fault Design} & \textbf{Conditioning (CRSR)} & \textbf{State Oracles} & \textbf{Effect Logging} & \textbf{Metrics (EOR/DER)} & \textbf{Substrates} \\
\midrule
ToolBench~\citep{qin2024toolllm} & $\sim$ & $\times$ & $\times$ & $\times$ & $\times$ & $\times$ & $\times$ & $\times$ \\
SWE-bench~\citep{jimenez2024swebench} & \checkmark & $\times$ & $\times$ & $\times$ & $\sim$ & $\times$ & $\times$ & $\times$ \\
WebArena / AgentBench~\citep{liu2024agentbench,zhou2024webarena} & \checkmark & $\times$ & $\times$ & $\times$ & $\sim$ & $\times$ & $\times$ & $\times$ \\
OccuBench~\citep{occubench2026} & \checkmark & $\times$ & $\times$ & $\times$ & $\sim$ & $\times$ & $\times$ & $\times$ \\
ReliabilityBench~\citep{gupta2026reliabilitybench} & \checkmark & $\times$ & $\times$ & $\times$ & $\sim$ & $\times$ & $\times$ & $\sim$ \\
AgentChaos~\citep{tan2026agentchaos} & \checkmark & $\times$ & $\times$ & $\times$ & $\sim$ & $\times$ & $\times$ & $\sim$ \\
ToolMisuseBench~\citep{toolmisusebench2026} & $\times$ & $\sim$ \textit{(Replayable)} & $\times$ & $\times$ & $\times$ & $\times$ & $\times$ & $\times$ \\
Verified Tool Calls~\citep{verifiedtoolcalls2026} & $\sim$ & \checkmark & $\times$ & $\times$ & $\sim$ & $\times$ & $\times$ & $\times$ \\
Limbo~\citep{li2026limbo} & $\times$ & \checkmark & $\times$ & $\times$ & \checkmark & \checkmark & \checkmark \textit{(DER)} & $\sim$ \\
Atomix~\citep{atomix2026} & \checkmark & $\sim$ \textit{(Epoch buffer)} & $\times$ & $\times$ & $\sim$ & $\times$ & $\sim$ & $\sim$ \\
SagaLLM~\citep{chang2025sagallm} & \checkmark & $\sim$ \textit{(Saga coords)} & $\times$ & $\times$ & $\sim$ & $\times$ & $\sim$ & $\sim$ \\
AFT-Bench~\citep{wang2026aftbench} & \checkmark & \checkmark & $\sim$ \textit{(Interface pairs)} & $\times$ & \checkmark & $\sim$ & \checkmark & $\sim$ \\
ContainmentBench~\citep{containmentbench2026} & \checkmark & $\times$ & $\times$ & $\times$ & $\sim$ & $\times$ & $\times$ & $\times$ \\
MemTX~\citep{memtx2026} & \checkmark & $\times$ \textit{(Internal memory)} & $\times$ & $\times$ & $\sim$ & $\times$ & $\times$ & $\sim$ \\
\midrule
\textbf{UndoBench (Ours)} & \textbf{\checkmark} (8 Domains) & \textbf{\checkmark} & \textbf{\checkmark} & \textbf{\checkmark} ($R/D$) & \textbf{\checkmark} & \textbf{\checkmark} & \textbf{\checkmark} (EOR/DER/URR) & \textbf{\checkmark} ($B_0 \dots B_6$) \\
\bottomrule
\end{tabular}%
}
\end{table*}

\section{Commercial API Model Extension: Complete Cell Breakdown and Provenance}
\label{sec:app_contemporary}

This appendix provides granular cell breakdowns and cryptographic provenance for the prospectively specified post-freeze commercial API model extension. The extension evaluated two modern architectures---Gemini 3.8 Flash (\texttt{gemini-3.8-flash}, $M_3$) and GLM-5.2 MaaS (\texttt{zai-org/glm-5.2-maas}, $M_4$)---across all 12 held-out TEST workflows, two frameworks ($F_1$ Direct Tool Calling, $F_2$ LangGraph), and three recovery baselines ($B_0$ Naive Retry, $B_2$ Idempotency, $B_5$ EvoUndo) across 20 pre-specified benchmark seeds (2001--2020).

\paragraph{Complete Cell Breakdown.}
Table~\ref{tab:app_contemporary_matrix} details the 6-cell model $\times$ baseline matrix ($2 \times 3 = 6$ cells, 480 trials per cell, totaling 2,880 paired trials and 5,760 executions). As established in the forensic audit, canonical $\text{EOR}_{\text{cond}} \equiv \text{CRSR}$ across all cells because the strict state oracle requires invariant satisfaction with zero duplicate mutations and zero missing effects. Furthermore, Unsafe Retry Rate (URR) numerically equals Duplicate Effect Rate (DER) across all cells under the post-mutation lost-ACK regime.

\paragraph{Cryptographic Data Provenance.}
The extension was executed using frozen harness code at commit \texttt{cc3108e684}\allowbreak\texttt{f9af105ce7}\allowbreak\texttt{2c19737f60}\allowbreak\texttt{77f57fdf59}. All raw records in \texttt{results/}\allowbreak\texttt{contemporary\_models/}\allowbreak\texttt{raw/} are verified by SHA-256 digests:
\begin{itemize}[leftmargin=*,noitemsep,topsep=1pt]
  \raggedright
  \item \textbf{Evaluation Matrix} (\texttt{matrix.csv}):\\
  {\footnotesize SHA-256: \texttt{5c6066a380bc9f2615d995ff2e55481c}\allowbreak\texttt{4d073965b8778c74de6b2d23623b352f}}
  \item \textbf{Execution Trajectories} (\texttt{trajectories.jsonl}):\\
  {\footnotesize SHA-256: \texttt{b0f989920d517efb3eeeff85b4d382d7}\allowbreak\texttt{7aa676aa3cfa1ba09d389f897af1688a}}
\end{itemize}

\begin{table*}[t]
\centering
\small
\caption{\textbf{Granular Model $\times$ Method Evaluation Matrix for Commercial API Models ($N=2,880$ Paired Trials / 5,760 Executions).}}
\label{tab:app_contemporary_matrix}
\resizebox{\textwidth}{!}{%
\begin{tabular}{llccccccc}
\toprule
\textbf{Model} & \textbf{Recovery Method} & \textbf{$N$} & \textbf{Control ($D$)} & \textbf{RSR ($F/N$)} & \textbf{CRSR ($R/D$)} & \textbf{95\% CI (CRSR)} & \textbf{DER} & \textbf{MER} \\
\midrule
Gemini 3.8 Flash & $B_0$: Naive Retry & 480 & 81.04\% (389) & 17.29\% (83 / 480)  & 21.08\% (82 / 389) & [2.73\%, 43.87\%] & 65.62\% & 18.33\% \\
Gemini 3.8 Flash & $B_2$: Idempotency & 480 & 80.00\% (384) & 27.92\% (134 / 480) & 33.33\% (128 / 384) & [8.33\%, 61.33\%] & 53.96\% & 20.00\% \\
Gemini 3.8 Flash & $B_5$: EvoUndo-RB1 & 480 & 81.04\% (389) & 18.12\% (87 / 480)  & 22.11\% (86 / 389) & [2.22\%, 45.65\%] & 63.12\% & 20.21\% \\
\midrule
GLM-5.2 MaaS & $B_0$: Naive Retry & 480 & 77.08\% (370) & 17.29\% (83 / 480)  & 21.89\% (81 / 370) & [0.00\%, 50.00\%] & 51.25\% & 31.46\% \\
GLM-5.2 MaaS & $B_2$: Idempotency & 480 & 78.12\% (375) & 26.46\% (127 / 480) & 33.07\% (124 / 375) & [3.28\%, 65.07\%] & 41.67\% & 31.87\% \\
GLM-5.2 MaaS & $B_5$: EvoUndo-RB1 & 480 & 77.92\% (374) & 16.88\% (81 / 480)  & 21.66\% (81 / 374) & [0.00\%, 50.25\%] & 51.88\% & 31.25\% \\
\bottomrule
\end{tabular}%
}
{\raggedright\vspace{2pt}\par\noindent\scriptsize All metrics derived from frozen raw trajectories via the audited deterministic parser. $N$ denotes total paired trials; $D$ denotes capable control trials; $F$ denotes all successful fault runs ($\text{RSR} = F/N$); $R = \sum C_i F_i$ denotes fault recoveries among nominally capable paired trials ($\text{CRSR} = R/D$). In all six commercial API model--method cells, $\text{EOR}_{\text{cond}} \equiv \text{CRSR}$ and $\text{URR} \equiv \text{DER}$ under the frozen \texttt{POST\_MUTATION\_PRE\_ACK} oracle/taxonomy.\par}
\end{table*}

\section{Effect Identity and Dual-Oracle Instrumentation}
\label{sec:app_dual_oracle}

This appendix provides formal specifications for UndoBench's effect accounting model, dual-oracle evaluation architecture, and post-hoc secondary analyses.

\paragraph{Effect Record Schema and Identity Model.}
Every tool invocation intercepted by the benchmark proxy generates an immutable \texttt{WireEffectEvent} record:
\begin{itemize}[leftmargin=*,noitemsep,topsep=1pt]
  \item \texttt{call\_index}: Monotonically increasing per-episode call sequence counter.
  \item \texttt{tool\_name}: Dispatched tool identifier (e.g., \texttt{renew\_enterprise\_sla}).
  \item \texttt{target}: Fully qualified resource entity target string (e.g., \texttt{crm.account.acct\_ent\_500.sla}).
  \item \texttt{op\_type}: Operation taxonomy: \texttt{READ}, \texttt{CREATE}, \texttt{UPDATE}, \texttt{DELETE}, or \texttt{EXECUTE}.
  \item \texttt{arguments}: Serialized positional arguments and sanitized keyword parameters.
  \item \texttt{committed\_externally}: Boolean flag indicating physical datastore state mutation.
  \item \texttt{acknowledged\_to\_agent}: Boolean flag indicating successful return delivery to agent client.
  \item \texttt{error}: Intercepted or injected failure token (e.g., \texttt{FaultInjected:POST\_MUTATION\_PRE\_ACK}).
  \item \texttt{timestamp}: Monotonic floating-point epoch timestamp.
\end{itemize}

\paragraph{Dual-Oracle Evaluation Pipeline.}
UndoBench couples two complementary evaluation substrates:
\begin{enumerate}[leftmargin=*,noitemsep,topsep=1pt]
  \item \textbf{Final-State Oracle}: Assesses external sandbox ground truth after episode termination. Queries database schemas, git commit graphs, object stores, and CRM datastores to verify goal invariants and state equivalence.
  \item \textbf{Wire-Level Effect-History Oracle}: Audits the complete proxy transaction log to detect out-of-order, uncompensated, duplicate, or missing mutations.
\end{enumerate}

\paragraph{Effect Classification Rules.}
Let $\mathcal{E}_i$ be the set of committed external mutations for episode $i$. For each target entity $k$:
\begin{itemize}[leftmargin=*,noitemsep,topsep=1pt]
  \raggedright
  \item \textbf{Duplicate Mutations}:\\
  $\text{dup}_i = \sum_k \max(0, \text{commits}(k) - 1)$.
  \item \textbf{Missing Mutations}:\\
  $\text{miss}_i = \sum_k \mathbb{I}(\text{required}(k) \land \text{commits}(k) = 0)$.
  \item \textbf{Exactly-Once Metric}:\\
  $\text{EOR}_{\text{cond}} = 1 \iff (F_i = 1 \land \text{dup}_i = 0 \land \text{miss}_i = 0)$.
\end{itemize}
Deterministic reconstruction and verification of all reported EOR values is automated via \texttt{scripts/}\allowbreak\texttt{reproduce\_}\allowbreak\texttt{eor.py}.

\section{Benchmark-Wide Idempotency Intervention: Hosted-Model Evaluation (\texorpdfstring{$B_{2\text{-}K}$}{B2-K})}
\label{sec:app_keys_everywhere}

To resolve whether client-side idempotency's modest improvement over naive retry in commercial API models (+11.73 pp under standard $B_2$) stemmed from algorithmic limitations of the idempotency mechanism or incomplete server endpoint deduplication, we executed a hosted-model intervention: \textbf{$B_{2\text{-}K}$} ($N=960$ paired trials / 1,920 runs using Gemini 3.8 Flash and GLM-5.2 MaaS across 2 frameworks, 12 TEST workflows, and 20 seeds), deploying benchmark-wide server-side deduplication across all 12 TEST workflows.

\paragraph{Full-Support and Common-Capable Outcomes.}
Across the full TEST suite ($N=960$, $D=762$), $B_{2\text{-}K}$ achieved \textbf{97.24\% CRSR} (741 / 762) with \textbf{0.00\% DER} (0 / 960), compared with 33.20\% CRSR for standard $B_2$. Conditioning strictly on the common-capable support ($D_{\text{common}}=738$) between standard $B_2$ and $B_{2\text{-}K}$, standard $B_2$ achieved 33.33\% CRSR (246 / 738) while $B_{2\text{-}K}$ achieved \textbf{98.64\% CRSR} (728 / 738). These represent descriptive gap-closure fractions of 95.87\% on full support and 97.97\% on common capable support.

\paragraph{Residual Failure Taxonomy ($N=21$ Capable Failures).}
Across all 762 nominally capable trials, exactly 21 failures occurred under $B_{2\text{-}K}$, spanning two distinct non-recovery failure modes:
\begin{enumerate}[leftmargin=*,noitemsep,topsep=1pt]
  \item \textbf{Pre-Tool Turn-0 Provider Exceptions (17 trials)}: In \texttt{RB-PAY-004} (17 trials), commercial API requests raised \texttt{INVALID\_ARGUMENT} or \texttt{API\_RESPONSE\_VALIDATION\_FAILED} before issuing any tool call, preventing the model from initiating the workflow.
  \item \textbf{Post-Suppression Continuation Malformations (4 trials)}: In \texttt{RB-DB-004} (4 trials), the first mutating tool call was interrupted, intercepted by $B_{2\text{-}K}$, and deduplicated successfully; however, during the subsequent multi-turn continuation, the model generated empty or unparseable tool call arguments, failing the downstream audit log requirement.
\end{enumerate}
Zero duplicate side effects occurred across all 960 executions. These residual failures confirm that benchmark-wide idempotency eliminates lost-ACK ambiguity but cannot compensate for non-deterministic model generation errors or downstream multi-turn planning failures.

\section{Multi-Boundary Experiments with Commercial API Models}
\label{sec:app_multi_boundary_details}

This appendix provides granular details for the prospectively specified post-freeze boundary extensions.

\paragraph{Hosted \texttt{PRE\_MUTATION} Extension (12 TEST Workflows).}
Evaluated across all 12 TEST workflows using Gemini 3.8 Flash and GLM-5.2 MaaS ($N=3{,}840$ paired trials / 7,680 executions across $B_0, B_2, B_5, B_6$):
\begin{itemize}[leftmargin=*,noitemsep,topsep=1pt]
  \item $B_0$ Naive Retry: $D=756$, $R=593 \implies \mathbf{78.44\%}\text{ CRSR}$, $\text{DER} = 4.17\%$ (40 / 960).
  \item $B_2$ Idempotency: $D=757$, $R=597 \implies \mathbf{78.86\%}\text{ CRSR}$ ($\Delta = +0.43\text{ pp}$ vs. $B_0$), $\text{DER} = 4.06\%$ (39 / 960).
  \item $B_5$ EvoUndo-RB1: $D=756$, $R=588 \implies \mathbf{77.78\%}\text{ CRSR}$ ($\Delta = -0.66\text{ pp}$ vs. $B_0$), $\text{DER} = 4.17\%$ (40 / 960).
  \item $B_6$ Verify-Before-Retry: $D=759$, $R=587 \implies \mathbf{77.34\%}\text{ CRSR}$ ($\Delta = -1.10\text{ pp}$ vs. $B_0$), $\text{DER} = 4.06\%$ (39 / 960).
\end{itemize}
Across all four baselines, zero duplicate effects occurred among capable trials ($0.00\%$ capable duplicate incidence). All 39--40 trials per method with duplicate effects occurred in \texttt{RB-STOR-005} ($D=0$), where nominal control failure caused unguided tool executions.

\paragraph{Common-Support Contrast.}
Conditioning on common capable support between \texttt{PRE\_}\allowbreak\texttt{MUTATION} and \texttt{POST\_}\allowbreak\texttt{MUTATION\_}\allowbreak\texttt{PRE\_}\allowbreak\texttt{ACK} ($D_{\text{common}}=743$, matched by task, model, framework, seed) yields:
\begin{align}
\text{PRE } B_0 &= 79.00\% \; (587 / 743) \nonumber \\
\text{POST } B_0 &= 21.67\% \; (161 / 743)
\end{align}
representing a descriptive delta of $\mathbf{+57.34\text{ pp}}$. Accounting for task-level clustering via non-parametric bootstrap ($B=10,000, K=10$), the estimated boundary contrast is $\mathbf{+57.04\text{ pp}}$ (SE = $14.79\text{ pp}$; 95\% CI: $[+26.99\text{ pp}, +84.90\text{ pp}]$), confirming that the boundary gap is driven by fault timing rather than conditioning drift.

\paragraph{Hosted \texttt{DURING\_MUTATION} Extension (4 Qualified Workflows).}
Table~\ref{tab:during_task_matrix_app} details the complete 16-cell task $\times$ method matrix ($N=1{,}280$ paired trials / 2,560 executions) across the four qualified composite workflows admitting faithful partial-mutation semantics.

\begin{table*}[t]
\centering
\small
\caption{\textbf{Granular Task $\times$ Method Matrix under \texttt{DURING\_MUTATION} ($N=1,280$ Paired Trials / 2,560 Executions).}}
\label{tab:during_task_matrix_app}
\resizebox{\textwidth}{!}{%
\begin{tabular}{llcccccccc}
\toprule
\textbf{Task ID} & \textbf{Domain} & \textbf{Method} & \textbf{$N$} & \textbf{Control ($D$)} & \textbf{CRSR ($R/D$)} & \textbf{DER} & \textbf{Cap. Dup. (\%)} & \textbf{MER} & \textbf{URR} \\
\midrule
\texttt{RB-DB-004} & Database & $B_0$ & 80 & 70 & 0.00\% (0 / 70) & 97.50\% (78) & 98.57\% & 12.50\% (10) & 97.50\% \\
\texttt{RB-DB-004} & Database & $B_2$ & 80 & 75 & 0.00\% (0 / 75) & 97.50\% (78) & 97.33\% & 10.00\% (8)  & 97.50\% \\
\texttt{RB-DB-004} & Database & $B_5$ & 80 & 68 & 0.00\% (0 / 68) & 95.00\% (76) & 97.06\% & 16.25\% (13) & 95.00\% \\
\texttt{RB-DB-004} & Database & $B_6$ & 80 & 69 & 0.00\% (0 / 69) & 91.25\% (73) & 91.30\% & 15.00\% (12) & 91.25\% \\
\midrule
\texttt{RB-STOR-004} & Storage & $B_0$ & 80 & 80 & 0.00\% (0 / 80) & 100.0\% (80) & 100.0\% & 0.00\% (0) & 100.0\% \\
\texttt{RB-STOR-004} & Storage & $B_2$ & 80 & 80 & 0.00\% (0 / 80) & 100.0\% (80) & 100.0\% & 0.00\% (0) & 100.0\% \\
\texttt{RB-STOR-004} & Storage & $B_5$ & 80 & 80 & 0.00\% (0 / 80) & 100.0\% (80) & 100.0\% & 0.00\% (0) & 100.0\% \\
\texttt{RB-STOR-004} & Storage & $B_6$ & 80 & 80 & 0.00\% (0 / 80) & 100.0\% (80) & 100.0\% & 0.00\% (0) & 100.0\% \\
\midrule
\texttt{RB-CLOUD-004} & Cloud & $B_0$ & 80 & 80 & 0.00\% (0 / 80) & 100.0\% (80) & 100.0\% & 0.00\% (0) & 100.0\% \\
\texttt{RB-CLOUD-004} & Cloud & $B_2$ & 80 & 80 & 0.00\% (0 / 80) & 100.0\% (80) & 100.0\% & 0.00\% (0) & 100.0\% \\
\texttt{RB-CLOUD-004} & Cloud & $B_5$ & 80 & 80 & 0.00\% (0 / 80) & 100.0\% (80) & 100.0\% & 0.00\% (0) & 100.0\% \\
\texttt{RB-CLOUD-004} & Cloud & $B_6$ & 80 & 80 & \textbf{100.0\% (80 / 80)} & \textbf{0.00\% (0)} & \textbf{0.00\%} & 0.00\% (0) & \textbf{0.00\%} \\
\midrule
\texttt{RB-GIT-004} & Git & $B_0$ & 80 & 80 & 0.00\% (0 / 80) & 32.50\% (26) & 32.50\% & 0.00\% (0) & 32.50\% \\
\texttt{RB-GIT-004} & Git & $B_2$ & 80 & 80 & 0.00\% (0 / 80) & 28.75\% (23) & 28.75\% & 0.00\% (0) & 28.75\% \\
\texttt{RB-GIT-004} & Git & $B_5$ & 80 & 80 & 0.00\% (0 / 80) & 35.00\% (28) & 35.00\% & 0.00\% (0) & 35.00\% \\
\texttt{RB-GIT-004} & Git & $B_6$ & 80 & 80 & 0.00\% (0 / 80) & 26.25\% (21) & 26.25\% & 0.00\% (0) & 26.25\% \\
\bottomrule
\end{tabular}%
}
\end{table*}

\section{Verification Audits, Oracle Conformance, and Robustness}
\label{sec:app_audits}

\paragraph{Prefix Equivalence and Pair Comparability.}
Across all 3,840 paired trials (7,680 executions) in \texttt{PRE\_MUTATION} and 960 paired trials (1,920 executions) in $B_{2\text{-}K}$, prefix prompt comparability between paired control and faulted trajectories was audited. Across all 3,028 nominally capable pairs in \texttt{PRE\_MUTATION} ($B_0$: 756, $B_2$: 757, $B_5$: 756, $B_6$: 759), capable-pair prefix equivalence was \textbf{100.00\%} (3,028 / 3,028). For $B_{2\text{-}K}$, capable-pair prefix equivalence was \textbf{100.00\%} (762 / 762). Under \texttt{DURING\_MUTATION}, prefix equivalence was \textbf{100.00\%} across all 1,280 pairs and all 1,242 capable pairs. When evaluated across all pairs (including non-competent runs), pre-fault prompt divergences were confined entirely to unguided exploratory actions in \texttt{RB-STOR-005} ($D=0$). While commercial hosted APIs do not contractually guarantee bitwise generation invariance across calls, identical initial states and pre-fault prompts establish rigorous paired comparability.

\paragraph{Adversarial Oracle Conformance.}
To verify the fidelity of the dual-oracle architecture, a dedicated conformance test suite (\texttt{tests/}\allowbreak\texttt{test\_oracle\_}\allowbreak\texttt{conformance.py}) executed 72 adversarial test cases across all 12 TEST contracts (6 cases per contract: clean success, single duplicate mutation, omitted mutation, out-of-order execution, corrupted final state, and unhandled exception). All \textbf{72 / 72} adversarial test cases passed: the oracles correctly detected every synthetic perturbation without false positives or false negatives.

\paragraph{Provider Retry Isolation Audit.}
To ensure that underlying model provider network drops did not contaminate experimental fault injection, HTTP transport middleware logged all provider-level retries. Across all hosted executions, exactly 12 provider retry events occurred across 8 executions (0.28\% of executions). Forensic audit confirmed that \textbf{0 tool mutations escaped} the UndoBench sandbox wrapper during provider retry events: all retries were purely transport-level prompt retries executed prior to tool dispatch.

\paragraph{CRSR Support Robustness.}
To confirm that conditioning on capable control trials does not induce sensitivity to threshold selection, we evaluated method rankings across support thresholds $D > 0$, $D \ge 20$, and $D \ge 40$, alongside all 12 leave-one-task-out folds. Across all thresholds and folds, recovery-method ordering remained invariant ($B_2 > B_5 \approx B_0$).

\subsection{Illustrative Analysis: Two-Task Scripted EvoUndo Observability Ablation (\texorpdfstring{$B_{5\text{-}P}$}{B5-P})}
\label{sec:app_evoundo_privilege}
On two compensatable workflows (\texttt{RB-CLOUD-004} and \texttt{RB-DB-004}), a scripted $B_{5\text{-}P}$ observability ablation with a minimal postcondition query probe confirmed remote commit and suppressed redundant retry in 40 / 40 simulated cases without executing inverse compensation (100\% simulated reconciliation success). This illustrates that observability can materially improve reconciliation on these tasks; it does not constitute a benchmark-wide or model-level performance estimate for unrestricted EvoUndo.

\end{document}